\documentclass[10pt,journal,compsoc]{IEEEtran}
\usepackage{graphicx}
\usepackage{subcaption}
\usepackage{amsmath}
\usepackage{amsthm}
\usepackage{multirow}
\usepackage{booktabs}
\usepackage{algorithm}
\usepackage{algorithmic}
\usepackage{bm}
\usepackage{enumitem}
\usepackage{xspace}
\usepackage{color}
\usepackage{soul, xcolor}

\newcommand{\ie}{\emph{i.e.,}\xspace}
\newcommand{\eg}{\emph{e.g.,}\xspace}
\newcommand{\et}{\emph{et al.}\xspace}

\usepackage{amssymb}
\def\B#1{\mathbf #1}
\def\C#1{\mathcal #1}

\ifCLASSOPTIONcompsoc
  \usepackage[nocompress]{cite}
\else
  \usepackage{cite}
\fi
\ifCLASSINFOpdf
\else
\fi
\begin{document}
%
\title{Learning Hierarchical Review Graph Representations for Recommendation}
%
%
%
%

\setstcolor{blue}

\author{
    Yong~Liu,
    Susen~Yang,
    Yinan~Zhang,
    Chunyan~Miao,
    Zaiqing~Nie,
    and~Juyong~Zhang

    \IEEEcompsocitemizethanks{
        \IEEEcompsocthanksitem Yong Liu is currently with Joint NTU-UBC Research Centre of Excellence in Active Living for the Elderly (LILY) and Alibaba-NTU Singapore Joint Research Institute, Nanyang Technological University, Singapore 639798. Email: stephenliu@ntu.edu.sg.
        \IEEEcompsocthanksitem Yinan Zhang and Chunyan Miao are currently with Joint NTU-UBC Research Centre of Excellence in Active Living for the Elderly (LILY), Alibaba-NTU Singapore Joint Research Institute, and School of Computer Science and Engineering, Nanyang Technological University, Singapore 639798. Email: yinan002@e.ntu.edu.sg, ascymiao@ntu.edu.sg.
        \IEEEcompsocthanksitem Susen Yang and Juyong Zhang are currently with School of Mathematical Sciences, University of Science and Technology of China, Hefei, Anhui, China 230052. Email: susen@mail.ustc.edu.cn, juyong@ustc.edu.cn.
        \IEEEcompsocthanksitem Zaiqing Nie is currently with the Alibaba Group, Beijing, China. Email: zaiqing.nzq@alibaba-inc.com.

    }

    \thanks{Manuscript received xxx, 2020; revised xxx, 2020.}
}

\IEEEtitleabstractindextext{%
\begin{abstract}
The user review data have been demonstrated to be effective in solving different recommendation problems. Previous review-based recommendation methods usually employ sophisticated compositional models, such as Recurrent Neural Networks (RNN) and Convolutional Neural Networks (CNN), to learn semantic representations from the review data for recommendation. However, these methods mainly capture the local dependency between neighbouring words in a word window, and they treat each review equally. Therefore, they may not be effective in capturing the global dependency between words, and tend to be easily biased by noise review information. In this paper, we propose a novel review-based recommendation model, named Review Graph Neural Network (RGNN). Specifically, RGNN builds a specific review graph for each individual user/item, which provides a global view about the user/item properties to help weaken the biases caused by noise review information. A type-aware graph attention mechanism is developed to learn semantic embeddings of words. Moreover, a personalized graph pooling operator is proposed to learn hierarchical representations of the review graph to form the semantic representation for each user/item. We compared RGNN with state-of-the-art review-based recommendation approaches on two real-world datasets. The experimental results indicate that RGNN consistently outperforms baseline methods, in terms of Mean Square Error (MSE).
\end{abstract}

}

\maketitle

\IEEEdisplaynontitleabstractindextext

%
\IEEEpeerreviewmaketitle

\IEEEraisesectionheading{\section{Introduction}\label{sec:introduction}}

\IEEEPARstart{W}{ith} the explosive growth of online information and contents, recommendation systems are playing an increasingly important role in various scenarios, \eg E-commerce websites and online social media platforms. In the last decades, various techniques~\cite{shi2014collaborative,zhang2019deep} have been proposed to solve different recommendation problems, for example business rating prediction~\cite{hu2014your}, Point-of-Interests recommendation~\cite{liu2013personalized}, and diversified item recommendation~\cite{wu2019pd}. Despite many research efforts have been devoted to developing recommendation systems, existing methods may still suffer from the data sparsity problem. To remedy this problem, various types of side information have been incorporated into the recommendation systems, 
\eg the item knowledge graph~\cite{sun2018recurrent}, and the users' reviews on items~\cite{chen2015recommender}. In this work, we focus on how to effectively exploit the user review data which usually describe users' preferences and also include their interests on different aspects of items.

Conventional review-based recommendation methods incorporate topic modeling techniques, \eg Latent Dirichlet Allocation (LDA)~\cite{blei2003latent} and word embedding model~\cite{mikolov2013efficient}, with matrix factorization to jointly learn semantic latent factors for users and items~\cite{mcauley2013hidden,bao2014topicmf,ling2014ratings,tan2016rating,zhang2016collaborative}. Recently, the rapid development of deep learning motivates the applications of different neural compositional models, \eg Convolutional Neural Network (CNN) and Recurrent Neural Network (RNN), to incorporate the review data for recommendation~\cite{wang2015collaborative,zheng2017joint,chen2018neural,liu2019daml,liu2019nrpa}. In addition, there also exists some recent work performing sentiment analysis on the review data to learn aspect-based representations for both users and items~\cite{bauman2017aspect,chin2018anr}. These review-based recommendation methods mainly focus on the rating prediction problem. Following these work, we also study the rating prediction task in this paper.


Despite the successes of existing review-based recommendation methods, they may still suffer from some of the following limitations. Firstly, the methods that only consider document-level co-occurrences of words~\cite{mcauley2013hidden,bao2014topicmf} can only capture the global-level semantic information. They usually ignore the word orders and the local context of words. Secondly, the word embedding and neural compositional models (\eg CNN and RNN) are effective in capturing the context information from neighbouring words in a word window. However, these methods treat each review equally. As they only rely on the local context of a word, they may be easily biased by the noise review information that are not relevant to the user interests or item properties.
Thirdly, most existing methods use the same model to learn the representations of all users and items, without considering the specific characteristics of an individual user/item.


To overcome above limitations, we propose a novel review-based recommendation method, namely Review Graph Neural Network (RGNN). The contributions made in this paper are summarized as follows.
\begin{itemize}
  \item We propose to build a directed review graph to describe the reviews of a user/item, where nodes represent the words and edges describe the co-occurrences and order relationships between words. The review graph aggregates the information of all the reviews associated with the user/item to provide a global view about the user/item properties. Then, the reviews with similar contents tend to have more dense connections with each other. Therefore, the review graph can help weaken the biases caused by the noise reviews.

  \item We develop the type-aware graph attention (TGAT) mechanism to learn the word embeddings, by aggregating the context information from neighboring words and considering the word orders.

  \item Moreover, we also develop the personalized graph pooling (PGP) operator to generate the hierarchical representations of the user/item review graph, considering the user/item-specific properties. PGP is able to select different informative words for different users/items, based on the intuition that the same word may have different semantic informativeness with respect to different users/items.

  \item To demonstrate the effectiveness of RGNN, we perform extensive experiments on two real datasets, compared with state-of-the-art review-based recommendation methods.

\end{itemize}

The remaining parts of this work are organized as follows. In Section~\ref{sec:relatedwork}, we review the most relevant existing work. Then, in Section~\ref{sec:model}, we introduce the details of the proposed recommendation model. Next, Section~\ref{sec:experiments} reports the experimental results. Finally, Section~\ref{sec:conclusion} concludes this work and discusses the potential future work directions.

\section{Related Work}
\label{sec:relatedwork}

This section reviews existing work about review-based recommendation, graph neural networks, and graph pooling.

\subsection{Review-based Recommendation Methods}
Traditional review-based recommendation methods employ topic modeling techniques on the review data to learn latent feature distributions of users and items. For example, McAuley and Leskovec~\cite{mcauley2013hidden} propose the HFT model that learns the latent factors of users and items from review texts by a LDA model. Guang \et~\cite{ling2014ratings} propose the RMR model that combines the topic model with a mixture of Gaussian models to improve the recommendation accuracy. In~\cite{bao2014topicmf}, Bao \et employ non-negative matrix factorization to derive topics of each review, and a transform function to align the topic distributions with corresponding latent user/item factors. In~\cite{tan2016rating}, Tan \et propose a rating-boosted method that combines textural reviews with users' sentiment orientations to learn more accurate topic distributions.

\begin{figure*}[t]
\centering
\includegraphics[width=0.75\textwidth]{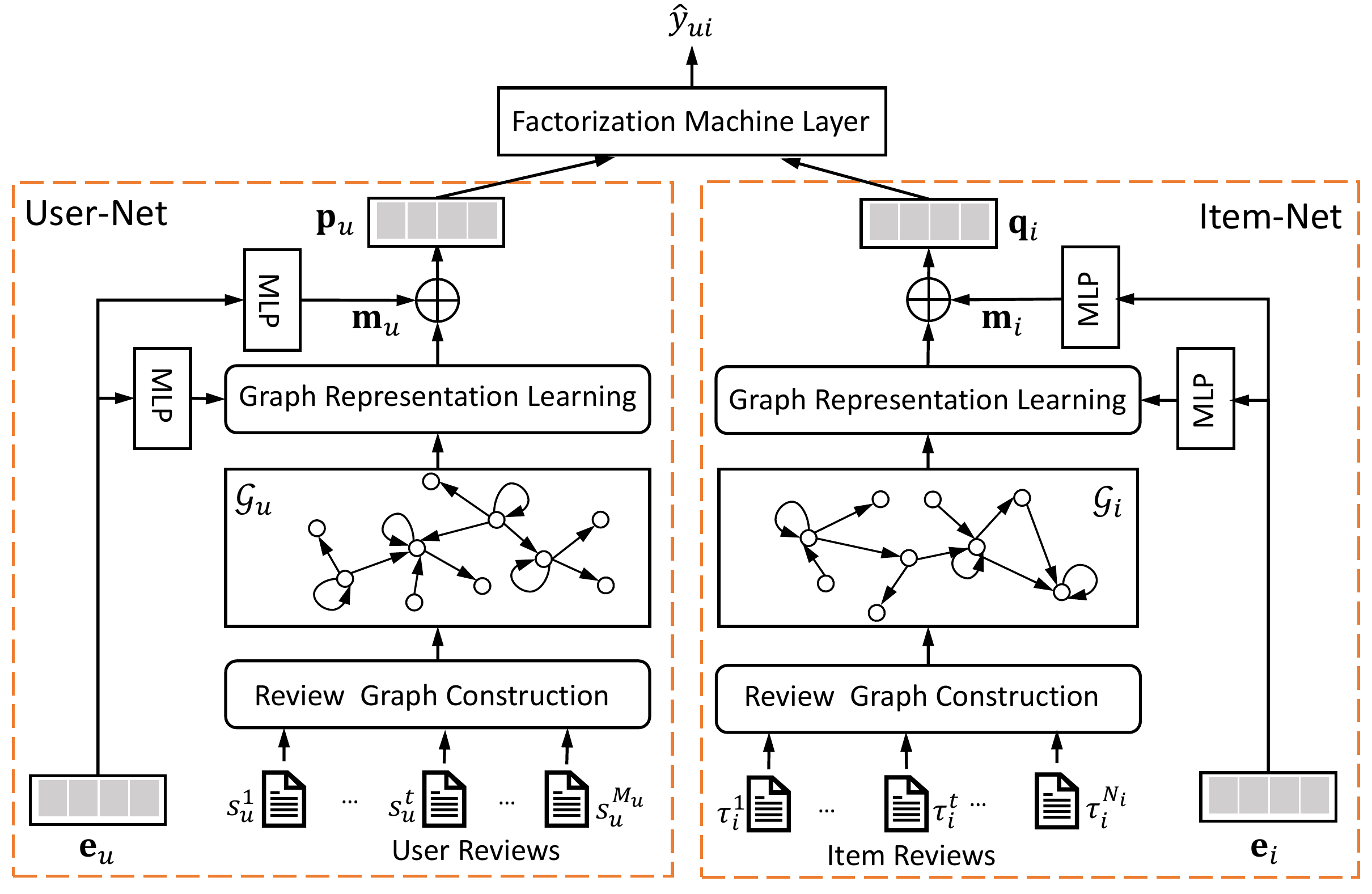}
\caption{The overall framework of the proposed RGNN model.}
\label{fig:framework}
\end{figure*}

The quick development of deep learning motivates the application of deep learning techniques to extract semantic information from the review data and build the end-to-end recommendation models. For instance, Wang \et~\cite{wang2015collaborative} propose the pioneer work that employs stacked denoising autoencoders (SDAE) to learn deep representations from review contents, and probabilistic matrix factorization to model users' rating behaviors. Zheng \et~\cite{zheng2017joint} propose the DeepCoNN model employing CNN to extract semantic information from the user reviews, and factorization machine to capture the interactions between the user and item features. Recently, the attention mechanism is applied to further improve the recommendation accuracy. In~\cite{chen2018neural}, Chen \et extract the review representation using CNN, and scored each review through an attention mechanism to learn informative user and item representations. In~\cite{wu2019context}, Wu \et utilize CNN with attention mechanism to highlight relevant semantic information by jointly considering the reviews written by a user and the reviews written to an item. Moreover, Lu \et~\cite{lu2018coevolutionary} propose to jointly learn the user and item information from ratings and reviews by optimizing matrix factorization and an attention-based GRU (\ie Gated Recurrent Unit) network. Liu \et~\cite{liu2019daml} propose the dual attention mutual learning between ratings and reviews for recommendation. They use both local and mutual attention of the CNN to jointly learn review representations for improving the model interpretability. Wu \et~\cite{wu2019reviews} utilize three attention networks to learn representations of sentence, review, and user/item in turn, and a graph attention network to model user-item interactions. To incorporate the user/item characteristics, Liu \et~\cite{liu2019nrpa} apply a personalized attention model to learn personalized representations of users and items from the review data. In addition, there are some aspect-based recommendation models~\cite{bauman2017aspect,chin2018anr} that apply aspect-based representation learning to automatically detect the most important parts of a review, and jointly learn the aspect, user, and item representations.

\subsection{Graph Neural Networks}

The graph neural networks (GNNs) extend deep learning techniques to process the graph data. Two main categories of GNN methods are recurrent graph neural networks and graph convolutional neural networks~\cite{wu2020comprehensive}.

The pioneer GNN methods are about recurrent graph neural networks, which learn the representation of a node by propagating its neighborhood information in an iterative manner until convergence. For example, in~\cite{scarselli2008graph}, the node's states are updated by exchanging its neighborhood information recurrently before satisfying the convergence criterion. Gallicchio and Micheli~\cite{gallicchio2010graph} improve the training efficiency of~\cite{scarselli2008graph} by generalizing the echo state network to graph domains. Moreover, Li \et~\cite{Li2015Gated} propose the gate graph neural networks (GGNN) which employs a gated recurrent unit as the recurrent function. This method no longer needs parameter constraints to ensure convergence.

The graph convolutional neural networks (GCNs) generalize the convolution operation from grid data to graph data. Existing GCN methods usually fall into two groups, \ie spectral-based methods and spatial-based methods. The spectral-based methods define convolution from the perspective of graph signal processing. For example, Bruna \et~\cite{Bruna2013Spectral} employs the eigen decomposition of the graph Laplacian and defines the convolution in the Fourier domain. To reduce computational complexity, Defferrard \et~\cite{deffer2016gcnn} approximate the convolutional filters by Chebyshey polynomials of the diagonal matrix of eigenvalues. Then, Kipf and Welling~\cite{kipf2016semi} develop a simpler convolutional architecture by the first-order approximation of the CheNet~\cite{deffer2016gcnn}. The spatial-based methods directly operate on the graph and propagate node information along edges. For instance, Micheli~\cite{micheli2009neural} proposes to directly sum up a node's neighborhood information and applies residual connection to memorize information over each layer. Atwood and Towsley~\cite{atwood2016dcnn} treat the graph convolution as a diffusion process, and they assume there is a certain transition probability from a node to its neighbor. As the number of neighbors of nodes varies, Hamilton \et~\cite{hamilton2017gsage} propose the GraphSage model which samples a fixed number of neighbors for each node and utilizes three different aggregators (\eg LSTM) to aggregate the neighbors' feature information. In~\cite{velivckovic2017graph}, Veli{\v{c}}kovi{\'c} \et propose the graph attention networks (GAT) that assumes the contributions of neighbor nodes to the central node are different. The multi-head attention mechanism is performed to calculate the importance of different neighbors. Moreover, Zhang \et~\cite{Zhang2018GaAN} employ a self-attention mechanism to compute an additional attention score for each attention head. In the recent work~\cite{keyu2019GIN}, Xu \et identify the graph structures that cannot be distinguished by popular GNN methods, and propose the Graph Isomorphism Network (GIN) which uses a learnable parameter to adjust the weight of the central node.

\subsection{Graph Pooling}

The pooling operation usually helps provide better performance in graph representation tasks. Earlier pooling methods are mainly based on the deterministic graph clustering algorithms~\cite{defferrard2016convolutional,fey2018splinecnn}, which are less flexible. The recent graph pooling methods can be divided into two groups: spectral and non-spectral methods. The spectral pooling methods~\cite{bianchi2019mincut,ma2019graph} employ laplace matrix and eigen-decomposition to capture the graph topology. However, these methods are not scalable to large graph, due to high computation complexity. Thus, a lot of research attentions have been attracted to develop the non-spectral pooling methods, which can be further divided into global and hierarchical pooling methods. The global pooling methods~\cite{vinyals2015order,zhang2018end} pool the representations of all nodes at one step. For example, Vinyals \et~\cite{vinyals2015order} propose the Set2Set model that aggregates nodes information by combining LSTM with a content-based attention mechanism. Zhang \et~\cite{zhang2018end} develop the SortPool model that first sorts nodes according to their features and then concatenates the embedding of sorted nodes to summarize the graph. Differing from the global pooling methods, the hierarchical pooling methods aim to learn hierarchical graph representations. For example, Ying \et~\cite{ying2018hierarchical} propose the first end-to-end differentiable pooling operator, named DiffPool, which learns a dense soft clustering assignment matrix of nodes by GNN. The gPool~\cite{gao2019graph,gao2019learning} and SAGPool~\cite{lee2019self} pooling operators utilize a node-scoring procedure to select a set of top ranked nodes to form an induced graph from original graph. Moreover, Diehl~\cite{diehl2019edge} proposes a hard pooling method called EdgePool, which is developed based on edge contraction. In~\cite{ranjan2019asap}, Ranjan \et utilize a self-attention mechanism to capture the importance of each node. This method learns a sparse soft cluster assignment for nodes to pool the graph. In~\cite{zhang2019hierarchical}, Zhang \et propose a non-parametric pooling operation using a structure learning mechanism with sparse attention to learn a refined graph structure that preserves the key substructure in original graph.

\section{The Proposed Recommendation Model}
\label{sec:model}

This work focuses on exploiting users' review data for recommendation. For each user $u$, the set of her reviews is denoted by $\C{S}_u = \{s_{u}^1, s_{u}^2, \cdots, s_{u}^{M_u}\}$, where $s_u^t$ denotes the $t$-th review of user $u$, and $M_{u}$ denotes the number of reviews generated by the user $u$. Similarly, let $\C{T}_i = \{\tau_{i}^1, \tau_{i}^2, \cdots, \tau_{i}^{N_i}\}$ denote the reviews of each item $i$, where $N_i$ denotes the number of reviews generated on the item $i$, and $\tau_{i}^t$ denotes the $t$-th review of item $i$. Note that each review may consist of multiple text sentences. In addition, we denote the observed rating given by the user $u$ on the item $i$ by $y_{ui}$. 
For a word $x$, we denote its word embedding by $\B{e}_x \in \mathbb{R}^{1 \times d_0}$, where $d_0$ denotes the dimensionality of the latent semantic space. Moreover, we use $\B{e}_u \in \mathbb{R}^{1 \times d_1}$ and $\B{e}_i \in \mathbb{R}^{1 \times d_1}$ to denote the embedding of user $u$ and item $i$, respectively, where $d_1$ is the dimensionality of latent rating space.

The overall framework of the proposed RGNN model is summarized in Figure~\ref{fig:framework}. As shown in Figure~\ref{fig:framework}, RGNN contains three components, \ie the User-Net, Item-Net, and a prediction layer. The User-Net and Item-Net are two parallel networks used to learn semantic representations for users and items, respectively. The prediction layer is developed based on factorization machine (FM)~\cite{rendle2010factorization}, which aims to predict the rating scores by considering the interactions between learned semantic features. Both User-Net and Item-Net consist of the following three modules:
\begin{itemize}
  \item The review graph construction module builds the review graph for each user and item, considering the review words as nodes in the graph and capturing the co-occurrence and order relationships between words.

  \item The graph representation learning module applies a type-aware graph attention network (TGAT) and personalized graph pooling operator (PGP) to extract the user/item-specific hierarchical representations from the entire review graph.

  \item Moreover, the feature fusion module integrates the hierarchical representations of the user/item review graph and the user/item embedding to form the semantic representation for the user/item.
\end{itemize}
As User-Net and Item-Net only differ in their inputs, we only introduce the details of User-Net in the following sections. The same procedure is applied to the Item-Net.

\subsection{Review Graph Construction}
\label{ss:graph}
Various graph based methods have been proposed to process the text data~\cite{malliaros2017graph}, which aim to capture the inherent dependence between words. Here, we utilize the graph-of-words method~\cite{mihalcea-tarau-2004-textrank} to construct the review graph with preserving the word orders. For the reviews $\C{S_u}$ of the user $u$, we firstly apply the phrase pre-processing techniques such as tokenization and text cleaning to select keywords of each review in $\C{S_u}$. Then, all reviews in $\C{S}_u$ can be encoded into an un-weighted and directed graph, where nodes represent words and edges describe the co-occurrences between words within a sliding window with fixed size $\omega$. Note that we define the word windows considering the positions of selected keywords in the original text.

Intuitively, the order information of words in the text may help capture the sentiment information~\cite{shen2018baseline}. For example, the phrases ``not really good'' and ``really not good'' convey different levels of negative sentiment information, which is only caused by the word orders. To preserve the orders of words in the review, we define three types of edges in the review graph: \textbf{forward type}, \textbf{backward type}, and \textbf{self-loop type} (respectively denoted by $e_f$, $e_b$, and $e_s$). We denote the set of edge type by $\C{R}=\{e_f, e_b, e_s\}$. In a review $s_u^t \in \C{S}_u$, if the selected keyword (\ie node) $x_2$ appears before the selected keyword $x_1$ and the distance between $x_1$ and $x_2$ in the original text of $s_u^t$ is \textbf{\emph{less than}} $\omega$, we construct an edge $E(x_2, x_1)$ from $x_2$ to $x_1$ and set the type of edge $E(x_2, x_1)$ as $e_f$. Moreover, we also construct an edge $E(x_1, x_2)$ from $x_1$ to $x_2$ and set the type of $E(x_1, x_2)$ as $e_b$. In the rare situation where the word $x_2$ may appear both before and after the word $x_1$, we randomly set the type of edge $E(x_1, x_2)$ as $e_f$ or $e_b$. In the experiments, we empirically set $\omega$ to 3.
To consider the information of the word itself (\eg $x_1$), we add a self-loop
edge at each word (\eg $E(x_1, x_1)$) and set the edge type as $e_s$. This trick also helps define the attention score in Eq.~\eqref{eq:attention}.

\begin{figure}
\centering
\includegraphics[width=0.75\columnwidth]{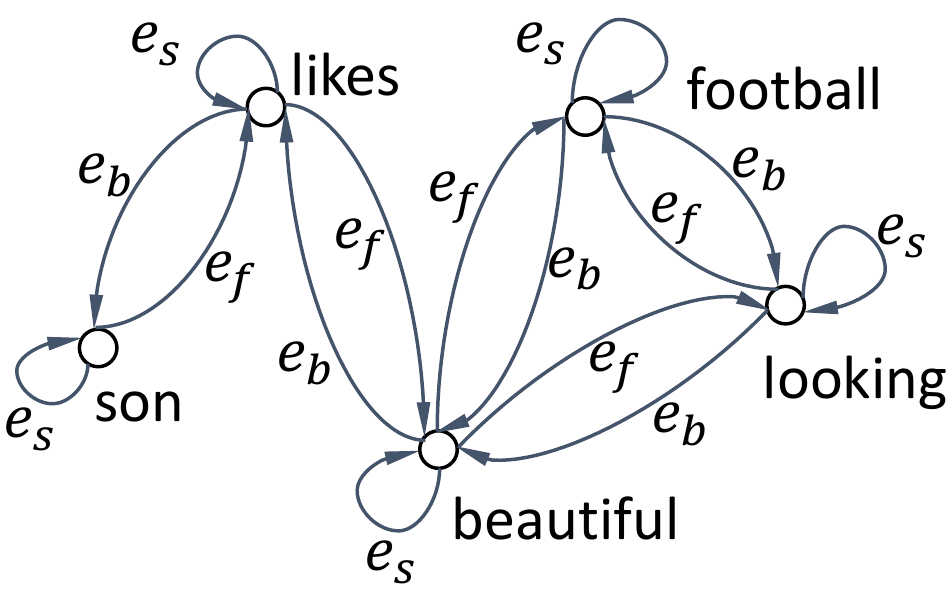}
\caption{An example of constructing graph for the review ``My son likes this beautiful looking football''. The words ``son'', ``likes'', ``beautiful'', ``looking'', and ``football'' are selected as nodes. Empirically, we set the window size $\omega$ to 3 for edge building.}
\label{fig:graphexam}
\end{figure}

Figure~\ref{fig:graphexam} shows an example about constructing the review graph for the review ``My \underline{son} \underline{likes} this \underline{beautiful} \underline{looking} \underline{football}'', where the words ``son'', ``likes'', ``beautiful'', ``looking'', and ``football'' are selected as keywords to build the review graph. In the original review text, the distance between ``son'' and ``likes'' is 1 which is less than $\omega$. Thus, we construct an edge from ``son'' to ``likes'' and set the edge type as $e_f$. Moreover, we also construct an edge from ``likes'' to ``son'' and set the edge type as $e_b$. However, the distance between ``son'' and ``beautiful'' in original text is 3, which is not smaller than the windows size ($\omega=3$). Therefore, as shown in Figure~\ref{fig:graphexam}, there is no connection between the selected keywords ``son'' and ``beautiful''.

For each user $u$, we denote her review graph by $\C{G}_u=\{\C{X}_u, \C{E}_u\}$, where $\C{X}_u$ denotes the set of nodes (\ie selected keywords) and $\C{E}_u$ denotes the set of node-edge-node triples $(x_h, e_r, x_t)$. Here, $e_r$ is type of the edge connecting from node $x_t$ to $x_h$, which is one of the three edge types defined above. Similarly, we can build the review graph $\C{G}_i=\{\C{X}_i, \C{E}_i\}$ for each item $i$.


\begin{figure*}
\centering
\includegraphics[width=0.95\textwidth]{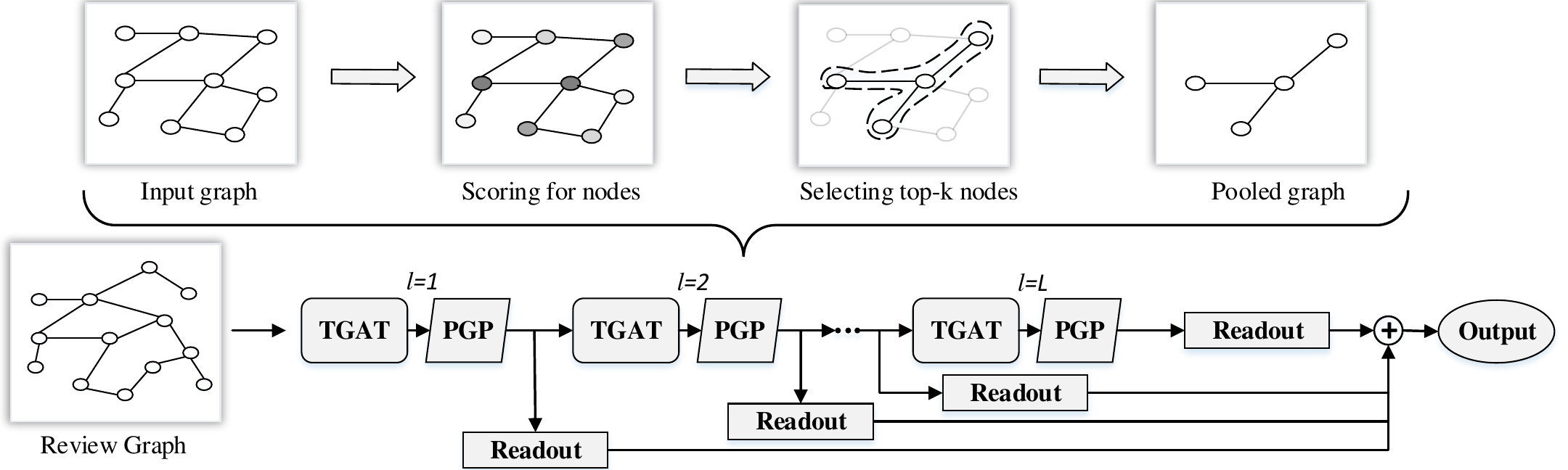}
\caption{The procedure of hierarchical graph representation learning.}
\label{fig:workflow}
\end{figure*}

\subsection{Hierarchical Graph Representation Learning}
After completing the construction of review graph $\C{G}_u$, we employ the representation learning procedure as shown in Figure~\ref{fig:workflow} to learn hierarchical representations of $\C{G}_u$, by repeating the TGAT and PGP operations multiple times. In this section, we introduce the details of the operations in the $\ell$-th representation learning layer, where the input graph is $\C{G}_u^{\ell}=\{\C{X}_u^{\ell}, \C{E}_u^{\ell}\}$. Suppose $\C{G}_u^{\ell}$ has $K_u^{\ell}$ nodes. Then, we also use an adjacency matrix $\B{A}^{\ell}_u \in \mathbb{R}^{K_u^{\ell} \times K_u^{\ell}}$ and a feature matrix $\B{X}_u^\ell \in \mathbb{R}^{K_u^{\ell} \times d_2^\ell}$ to describe $\C{G}_u^{\ell}$, where each row in $\B{X}_u^\ell(x_1,:)$ denotes the input feature of the node $x_1$, and $d_2^{\ell}$ denotes the dimensionality of the input feature. Node that the features of nodes in the review graph are initialized by the embeddings of corresponding words. Thus, at the first layer, $d_2^{\ell}$ is equal to $d_0$. In $\C{G}_u^\ell$, if there exists an edge between two nodes $x_1$ and $x_2$, we set $\B{A}^{\ell}_u[x_1, x_2]$ to 1, and set to 0, otherwise.

\subsubsection{Type-aware Graph Attention Network}
In the reviews, the semantic meaning of a word can be enriched by its neighboring words, which can be considered as the dependency between a node and its neighboring nodes in the review graph. At the $\ell$-th representation learning layer, for a node $x_h$ in the graph $\C{G}_u^\ell$, we denote its neighborhood in the graph by $\C{N}_h^{\ell}= \{x_t | (x_h, e_r, x_t) \in \C{E}_u^{\ell}\}$. Note that $\C{N}_h^{\ell}$ also includes the node $x_h$, because there exists an self-loop edge at each node in the review graph. A type-aware attention mechanism is developed to effectively aggregate the information of neighboring words, considering the types of connecting edges. Following~\cite{velivckovic2017graph}, we define the attention weight $a^{\ell}(x_h, e_r, x_t)$ of a neighboring node $x_t \in \C{N}_h^{\ell}$ as,
\begin{equation}
a^{\ell}(x_h, e_r, x_t) = \frac{\exp[\pi^{\ell}(x_h, e_r, x_t)]}{\sum_{x_{\tilde{t}} \in \C{N}_h^{\ell}}\exp[\pi^{\ell}(x_h, e_{\tilde{r}}, x_{\tilde{t}})]},
\label{eq:attention}
\end{equation}
where $\pi^{\ell}(x_h, e_r, x_t)$ is implemented by the following relational attention mechanism,
\begin{equation}
\pi^{\ell}(x_h,e_r,x_t) = \mbox{LeakyReLU}\big[(\B{x}^\ell_h\B{W}^\ell_1)(\B{x}^\ell_t\B{W}^\ell_2 + \B{e}^\ell_r\B{W}^\ell_3)^\top\big].
\label{eq:attentionscores}
\end{equation}
Here, $\B{x}_h^\ell, \B{x}_t^\ell \in \mathbb{R}^{1 \times d_2^\ell}$ are the input embeddings of $x_h$ and $x_t$ at the $\ell$-th layer. $\B{e}_r^\ell \in \mathbb{R}^{1 \times d_2^{\ell}}$ is the embeddings of $e_r$ at the $\ell$-th layer. $\B{W}_1^\ell, \B{W}_2^\ell, \B{W}_3^\ell \in \mathbb{R}^{d_2^{\ell} \times d_2}$ are three different weight matrices used for the linear transformations of the node and edge embeddings, and $d_2$ denotes the dimensionality of the hidden space of the graph representation learning layers.
The attention weights indicate which neighboring node should contribute more to the central node, when aggregating neighborhood information. Then, we can compute the linear combination of neighbors as follows,
\begin{equation}
\bar{\B{x}}_h^{\ell} = \sum_{x_t \in \C{N}_h^\ell}a^\ell(x_h, e_r, x_t)\B{x}^\ell_t\B{W}_4^\ell,
\label{eq:attentionvec}
\end{equation}
where $\B{W}_4^\ell \in \mathbb{R}^{d_2^\ell \times d_2}$ denotes the transform matrix. A feature matrix $\bar{\B{X}}_u^\ell \in \mathbb{R}^{K_u^{\ell} \times d_2}$ is used to denote the output of TGAT. Each row of $\bar{\B{X}}_u^\ell$ corresponds to the feature of a node $x$ in $\C{G}_u^\ell$, computed by Eq.~\eqref{eq:attentionvec}. Then,  $\bar{\B{X}}_u^\ell$ will be used as the input of the PGP operation at the $\ell$-th representation learning layer.

Compared with the methods using CNN to extract semantic information, the proposed TGAT operation can not only capture the global contextual word information from all the reviews of a user/item, but also specify varying importance weights to different contextual words. In addition, we explicitly incorporate the edge type $e_r$ into the attention mechanism, which encodes word order information during aggregating neighboring contextual word information.

\subsubsection{Personalized Graph Pooling}
The graph pooling operation plays an important role in learning graph representations. Various hierarchical graph pooling methods have been proposed to adaptively select a subset of important nodes to form a smaller sub-graph~\cite{gao2019graph,gao2019learning,lee2019self,ranjan2019asap,zhang2019hierarchical}. However, these methods ignore the user/item-specific properties of an individual user/item, thus they are not suitable to learn the review graph representation for recommendation. In practice, different users may emphasize different aspects of items. For example, some users are most interested in the item price, while some other users may be more interested in the item quality. Similarly, the most important properties of different items may also be different. Therefore, it is intuitively that the same word may have different importance to different users/items. Motivated by this intuition, we propose the personalized graph pooling operator to capture the important nodes (\ie words) for each user/item.

For the user $u$, we first use a Multilayer Perceptron (MLP) to generate her personalized measurement vector $\bm{\theta}_u^l$ for measuring the importance of nodes at the $\ell$-th representation learning layer,
\begin{equation}
\bm{\theta}_u^{\ell} = \mbox{ReLU}\big(\B{e}_u\B{W}_5^{\ell}+\B{b}_5^{\ell}\big),
\end{equation}
where $\B{W}_5^{\ell} \in \mathbb{R}^{d_1 \times d_2}$ and $\B{b}_5^{\ell} \in \mathbb{R}^{1 \times d_2}$ are the weight matrix and bias term of MLP at the $\ell$-th pooling layer.
We adapt graph pooling to the recommendation scenario studied in this work, and define the following importance score $\beta_h^\ell$ for each node $x_h \in \C{X}_u^\ell$ as,
\begin{equation}
\beta_h^{\ell} = \big|\bar{\B{x}}_h^{\ell}{\bm{\theta}_u^{\ell}}^{\top}\big| + \frac{1}{|\C{N}^{\ell}_h|} \sum_{x_t \in \C{N}^{\ell}_h} \big|(\bar{\B{x}}^{\ell}_h - \bar{\B{x}}^{\ell}_t) {\B{w}^{\ell}}^\top\big|,
\label{eq:infoscore}
\end{equation}
where $\B{w}^{\ell} \in \mathbb{R}^{1\times d_2}$ is a learnable parameter. In Eq.~\eqref{eq:infoscore}, the first part captures the importance of node $x_h$ to the user $u$. The second part describes the feature diversity of $x_h$'s neighborhood at the $\ell$-th layer. It is developed based on the criterion that a node $x_{h}$ can be removed in the pooled graph with negligible information loss, if $x_{h}$ can be well reconstructed by its neighboring nodes in $\C{G}_u^\ell$. Moreover, $\B{w}^{\ell}$ can also help balance the contributions of these two parts in the $\ell$-th layer.

Then, the score vector $\bm{\beta}^{\ell}=[\beta_1^{\ell},\beta_2^{\ell},\cdots,\beta_{K_u^{\ell}}^{\ell}]^\top$ is multiplied with the feature matrix $\bar{\B{X}}_u^l$ to obtain the following filtered feature matrix,
\begin{equation}
\widehat{\B{X}}_u^{\ell} = \B{D}^{\ell}\bar{\B{X}}_u^{\ell},
\end{equation}
where $\B{D}^{\ell} \in \mathbb{R}^{K_u^\ell \times K_u^\ell}$ is a diagonal matrix. The diagonal element $D_{hh}^\ell=\sigma(\beta_h^{\ell})$, and $\sigma(\cdot)$ denotes the sigmoid function. Then, we can obtain the adjacency matrix $\B{A}_u^{\ell+1}$ and feature matrix $\B{X}_u^{\ell+1}$ of the pooled graph $\C{G}_u^{\ell+1}$ as follows,
\begin{align}
&\mbox{idx} = \mbox{rank}(\bm{\beta}^{\ell}, K_u^{\ell+1}), \\ \B{A}_u^{\ell+1}=\B{A}_u^{\ell}&(\mbox{idx},\mbox{idx}),
\B{X}_u^{\ell+1} = \widehat{\B{X}}^{\ell}_u(\mbox{idx},:),
\end{align}
where $\mbox{rank}(\cdot)$ returns the indices of the top-$K_u^{\ell+1}$ values in $\bm{\beta}^{\ell}$, $\B{A}^{\ell}_u(\mbox{idx},\mbox{idx})$ is the row-wise and column-wise indexed adjacency matrix, and $\widehat{\B{X}}^{\ell}_u(\mbox{idx},:)$ is the row-wise indexed feature matrix. Then, we have $\B{A}_u^{\ell+1}$ and $\B{X}_u^{\ell+1}$ which are new adjacency matrix and corresponding feature matrix of a smaller graph $\C{G}_u^{\ell+1}$. In this work, we empirically set $K_u^{\ell+1}=\lceil\alpha K_{u}^{\ell}\rceil$, where $\alpha$ is the pooling ratio, $\lceil \cdot \rceil$ is the ceiling function.

\subsection{Feature Fusion}
After repeating the TGAT and PGP operations $L$ times, we can obtain multiple pooled sub-graphs with different size, \ie $\C{G}_u^2, \C{G}_u^3,\cdots, \C{G}_u^{L+1}$. This hierarchical pooling operation extracts the coarser-grained information of review graph layer-by-layer, which can be considered as different level representations of reviews, \eg word-level, sentence-level, and review-level. To integrate these sub-graph information, we first use the max-pooling and MLP to generate representation of each sub-graph $\C{G}_u^{\ell}$, where $2 \leq \ell \leq L+1$, as follows:
\begin{align}
\hat{\B{g}}_u^{\ell} = \mathop{\mbox{max}}\limits_{t=1}^{d_2} \B{X}_u^{\ell}(:,t),~~
\B{g}_u^{\ell} = \mbox{ReLU}(\hat{\B{g}}_u^{\ell}\B{W}_6^{\ell} + \B{b}_6^{\ell}),
\end{align}
where $\B{W}_6^{\ell} \in \mathbb{R}^{d_2 \times d_1}$ and $\B{b}_6^{\ell} \in \mathbb{R}^{1 \times d_1}$ are the weight matrix and bias vector. $\B{g}_u^{\ell}$ denotes the representation of the sub-graph generated at the $\ell$-th layer. Then, we concatenate the non-linear transform of the user embedding after a MLP layer and the representations of sub-graphs at all layers to generate the semantic representation $\B{p}_u$ of the user $u$ as follows,
\begin{align}
\B{m}_u &= \mbox{ReLU}(\B{e}_u\B{W}_7+\B{b}_7),\nonumber\\
\B{p}_u &= \B{m}_u \oplus \B{g}_u^2 \oplus \B{g}_u^3 \oplus \cdots \oplus \B{g}_u^{L+1},
\label{eq:userSem}
\end{align}
where $\oplus$ is the concatenating operation, $\B{W}_7 \in \mathbb{R}^{d_1 \times d_1}$ and $\B{b}_7 \in \mathbb{R}^{1 \times d_1}$ are the weight matrix and bias vector.
Similarly, we can get the semantic representation of the item $i$ as $\B{q}_i$.

\subsection{Rating Prediction}
A factorization machine (FM) layer is used to predict the user's preference on an item, considering the higher order interactions between the user and item semantic features. Specifically, we first concatenate the user and item semantic representations as $\B{z}=\B{p}_u \oplus \B{q}_i$, and the prediction $\hat{y}_{ui}$ of the user's preference can be defined as follows,
\begin{equation}
\hat{y}_{ui} = b_0 +b_u+ b_i + \B{z}\B{w}^\top + \sum_{i=1}^{d'}\sum_{j=i+1}^{d'}\langle\B{v}_i,\B{v}_j\rangle z_i z_j,
\label{eq:fmlayer}
\end{equation}
where $b_0$, $b_u$ and $b_i$ are the global bias, user bias, and item bias. $\B{w} \in \mathbb{R}^{1\times d'}$ is the coefficient vector, and $d'=2*(L+1)*d_1$. $\B{v}_i, \B{v}_j\in \mathbb{R}^{1 \times k}$ are the latent factors associated with $i$-th and $j$-th dimension of $\B{z}$. Empirically, we set $k=d_2$ in this work. $z_i$ is the value of the $i$-th dimension of $\B{z}$. In this work, we focus on the rating prediction problem. Thus, the model parameters $\B{\Theta}$ of RGNN can be learned by solving the following optimization problem,
\begin{equation}
 \min_{\B{\Theta}}\sum_{(u,i) \in \C{D}}\big(y_{ui}- \hat{y}_{ui}\big)^2 + \lambda\big\|\B{\Theta}\big\|_F^2,
\end{equation}
where $\lambda$ is the regularization parameter, $\C{D}$ denotes the set of user-item pairs used to update the model parameters. The entire framework can be effectively trained by the end-to-end back propagation algorithm.

\subsection{Discussions}

In this work, we build a review graph to describe all the reviews of a user/item. If a word appears in multiple review sentences, its neighbors in review graph aggregate all its neighboring words (within a word window) in these sentences. Then, TGAT is proposed to learn the word embedding, considering its global context words among all the relevant reviews. However, the CNN/RNN-based methods separately aggregate the local neighboring words of a word in different review sentences. Moreover, TGAT employs attention mechanism to automatically assign different importance weights to neighboring words. By performing multiple times of PGP operations, the proposed model can capture the long-term dependency between selected words.

Suppose the average number of selected keywords for each user/item is $N_w$. The time complexity of constructing the review graph is $O\big(\omega N_w\big)$. In practice, the review graph construction can be performed offline. The time complexity for RGNN to obtain the hierarchical review representation is $O\big(\sum_{\ell=1}^{L}\alpha^{\ell}N_wd_0^2\big)$, where $L$ denotes the number of layers, $\alpha$ and $d_0$ denote the pooling ratio and word embedding dimension. The total training time complexity of RGNN is $O\big(LN_wd_0^2\big)$. Moreover, the time complexity of using CNNs to obtain the review representation is $O\big(MSN_wd_0\big)$, where $M$ and $S$ denote the number of convolutional kernels and the convolutional kernel size, respectively. Thus, compared with the CNN based methods, the proposed RGNN model is not complicate or can be even more efficient.

\section{Experiments}
\label{sec:experiments}

We empirically perform extensive experiments on real datasets to demonstrate the effectiveness of the proposed RGNN method, comparing with state-of-the-art review-based recommendation methods.

\subsection{Experimental Setting}

\subsubsection{Datasets}
The experiments are performed on the Amazon review dataset~\cite{he2016ups} and Yelp dataset\footnote{https://www.yelp.com/dataset}, which have been widely used for recommendation research. For Amazon review dataset, we choose the following 5-core review subsets for evaluation: ``Musical Instruments'', ``Office Products'', ``Toys and Games'', ``Beauty'', and ``Video Games'' (respectively denoted by Music, Office, Toys, Beauty, and Games). For Yelp dataset, we only keep the users and items that have at least 10 reviews for experiments. Table~\ref{tab:dataset} summarizes the details of these experimental datasets.

\begin{table}
    \centering
    \small
    \caption{The statistics of the experimental datasets.}
    \label{tab:dataset}
    \begin{tabular}{l|c|c|c|c} \hline
    Dataset& \# Users & \# Items & \# Ratings & Density \\\hline
    Music  & 1,429 & 900 & 10,261 & 0.80\% \\
    Office & 4,905 & 2,420 & 53,228 & 0.45\%\\
    Toys & 19,412 & 11,924 & 167,597 & 0.07\%\\
    Beauty & 22,363 & 12,101 & 198,502 & 0.07\% \\
    Games  & 24,303 & 10,672 & 231,577 & 0.09\%\\
    Yelp & 126,084 & 65,786 & 3,519,533 & 0.04\% \\
    \hline
    \end{tabular}
\end{table}

\begin{table*}[]
\centering
\caption{The performances of different recommendation algorithms evaluated by MSE. The best results are in \textbf{bold} faces and the second best results are \underline{underlined}. $\blacktriangle$ denotes the average relative improvements achieved by RGNN over baselines on all datasets. $^\ast$ indicates RGNN significantly outperforms all baselines with $p < 0.05$ using student \emph{t-test} on the dataset.}
\begin{tabular}{l|cccccccccccc}
\hline
Dataset & PMF   & SVD++  & CDL    & DeepCoNN  & ANR & NARRE  & CARL & NRPA   & RMG    & DAML   & RGNN \\ \hline
Music   &1.8783 & 0.7952 & 1.2987 & 0.7909 & 0.7825 & 0.7688 &  0.7632 & 0.7658 & 0.7560 & \underline{0.7401} & \textbf{0.7319}$^\ast$ \\ \hline
Office  &0.9635 & 0.7213 & 0.8763 & 0.7315 & 0.7237 & 0.7266 &  0.7193 & 0.7343 & 0.7348 & \underline{0.7164} & \textbf{0.7125}$^\ast$ \\ \hline
Toys    &1.6091 & 0.8276 & 1.2479 & 0.8073 & 0.7974 & 0.7912 & 0.8248 & 0.8100 & 0.8435 & \underline{0.7909} & \textbf{0.7786}$^\ast$ \\\hline
Games   &1.5260 & 1.2081 & 1.6002 & 1.1234 & \underline{1.1038} & 1.1120 &  1.1308 & 1.1259 & 1.1496 & 1.1086 & \textbf{1.0996}$^\ast$ \\ \hline
Beauty  &2.7077 & 1.2129 & 1.7726 & 1.2210 & 1.2021 & \underline{1.1997} & 1.2250 & 1.2034 & 1.2419 & 1.2175 & \textbf{1.1885}$^\ast$ \\ \hline
Yelp    &1.4217	& 1.2973 & 1.4042 & 1.2719 & 1.2708	& \underline{1.2675} & 1.3199 & 1.2721 & 1.2981 & 1.2700 & \textbf{1.2645}$^\ast$ \\\hline
$\blacktriangle$ Improvements & 38.97\% & 4.77\% & 29.02\% & 3.16\% & 2.06\% & 1.77\% & 3.43\% & 2.57\% & 4.19\% & 1.14\% & -\\
\hline
\end{tabular}
\label{tab:results}
\end{table*}

\subsubsection{Setup and Metric}

For each dataset, we randomly choose 20\% of the user-item review pairs (denoted by $\C{D}_{test}$) for evaluating the model performance in the testing phase, and the remaining 80\% of the review pairs (denoted by $\C{D}_{train}$) are used in the training phase. Then, 10\% of the data in $\C{D}_{train}$ are held out as the validation set to choose the model hyper-parameters, and the remaining 90\% of the data in $\C{D}_{train}$ are used to update the model parameters.
As this work focuses on the rating prediction task, the performances of the recommendation algorithms are typically evaluated by Mean Square Error (MSE), which has been widely used in previous studies~\cite{zheng2017joint,chin2018anr,wu2019context,liu2019nrpa}. The definition of MSE is as follows:
\begin{align}
MSE=\frac{1}{|\mathcal{D}_{test}|}\sum_{(u, i)\in \mathcal{D}_{test}}(y_{ui}-\hat{y}_{ui})^{2},
\label{eq:mse}
\end{align}
where $\mathcal{D}_{test}$ denotes the set of all tested user-item rating pairs. From the definition in Eq.~\ref{eq:mse}, we can note that lower MSE indicates better performances.

\subsubsection{Baseline Methods}
We compare RGNN with the following recommendation methods:

\begin{itemize}
  \item \textbf{PMF}~\cite{mnih2008probabilistic}: This is the probabilistic matrix factorization model, which is a classical collaborative filtering based rating prediction method.

  \item \textbf{SVD++}~\cite{koren2008factorization}: This is a classic matrix factorization method that exploits both the user's explicit preferences on items and the influences of the user's historical items on the target item.

  \item \textbf{CDL}~\cite{wang2015collaborative}: This is a hierarchical Bayesian model that employs SDAE for learning features from the content information and collaborative filtering for modeling the rating behaviors.

  \item \textbf{DeepCoNN}~\cite{zheng2017joint}: This method contains two parallel networks, which focus on modeling the user behaviors and learning the item properties from the review data.

  \item \textbf{ANR}~\cite{chin2018anr}: This is aspect-based neural recommendation model that learns aspect-based representations for the user and item by an attention-based module. Moreover, the co-attention mechanism is applied to the user and item importance at the aspect-level.

  \item \textbf{NARRE}~\cite{chen2018neural}: This method uses an attention mechanism to model the importance of reviews and a neural regression model with review-level explanation for rating prediction.

  \item \textbf{CARL}~\cite{wu2019context}: This is a context-aware representation learning model for rating prediction, which uses convolution operation and attention mechanism for review-based feature learning and factorization machine for modeling high-order feature interactions.

  \item \textbf{NRPA}~\cite{liu2019nrpa}: This method uses a review encoder to learn the review representation, and a user/item encoder with personalized attention mechanism to learn user/item representations from reviews.

  \item \textbf{RMG}~\cite{wu2019reviews}: This method utilizes a multi-view learning framework to incorporate information from the review-content view and user-item interaction view for recommendation.

  \item \textbf{DAML}~\cite{liu2019daml}: This method employs CNN with local and mutual attention mechanism to learn the review features and improve the interpretability of the recommendation model.

\end{itemize}

\begin{table}[]
\centering
\caption{Performances of RGNN with respect to different settings of $L$.}
\begin{tabular}{l|cccccccc}
\hline
    & Music  & Office & Toys & Games & Beauty  & Yelp\\ \hline
L=1 & 0.7572 & 0.7130 & 0.7830 & 1.1021 & 1.1964 & 1.2671    \\ \hline
L=2 & \textbf{0.7319} & \textbf{0.7125} & \textbf{0.7786} & 1.1025 & 1.1946 & 1.2658\\ \hline
L=3 & 0.7557 & 0.7145 & 0.7825 & \textbf{1.0996} & 1.1903 & \textbf{1.2645}\\ \hline
L=4 & 0.7557 & 0.7149 & 0.7802 & 1.1017 & \textbf{1.1885} & 1.2684\\
\hline
\end{tabular}
\label{tab:LResults}
\end{table}

\subsubsection{Parameter Settings}
In this work, we implement the proposed RGNN model by Pytorch and choose Adam~\cite{kingma2014adam} as the optimization method to learn the model parameters. The dimensionality of the semantic space $d_0$, the rating space $d_1$, and the hidden space of the graph representation learning layers $d_2$ are chosen from $[4, 8, 16, 32, 48, 64, 128]$. The learning rate $\gamma$ is varied in $[0.0005, 0.001, 0.005]$, and regularization parameter $\lambda$ is varied in $[0.001, 0.01, 0.05, 0.1, 1.0]$. The batch size is chosen in $[64, 128, 256]$. Moreover, the number of layers $L$ is varied in $[1,2,3,4]$, and the pooling ratio $\alpha$ is set to 0.5.

The hyper-parameters of baseline methods are set following original papers. In PMF and SVD++, the dimension of latent embeddings are chosen from $\{16, 32, 48, 64, 128\}$. The learning rate is varied in $\{0.0001, 0.001, 0.01\}$, and the batch size is chosen from $\{64, 128, 256, 512\}$. For CDL, we set a=1, b=0.01, K=50, and choose ${\lambda}_u$, ${\lambda}_v$, ${\lambda}_n$, ${\lambda}_w$ from $\{0.01, 0.1, 1, 10, 100\}$. The number of hidden factors is selected from $\{50, 100, 200, 300\}$, and dropout ratio is varied in $\{0.1, 0.3, 0.5, 0.7, 0.9\}$. For DeepCoNN, we empirically vary the number of latent factors (\ie $|\B{x}_u|$ and $|\B{y}_i|$) from 5 to 100 and the number of convolutional kernels from 10 to 400. The window size is set to 3, and the learning rate is chosen from $\{0.0001, 0.001, 0.002, 0.01\}$. In ANR, the dimensionality of latent representations is chosen from $\{64, 128, 256\}$. The window size is set to 3. The number of aspect is selected from $\{2, 4, 6, 8\}$. Moreover, the dimension of aspect embeddings is chosen from $\{5, 10, 15, 25, 50\}$, and the dropout ratio is selected from $\{0.1, 0.3, 0.5, 0.7, 0.9\}$. For NARRE, the learning rate is selected from $\{0.0001, 0.001, 0.002, 0.01, 0.05\}$. The dropout ratio is chosen from $\{0.1, 0.3, 0.5, 0.7, 0.9\}$. The batch size is selected from $\{50, 100, 150\}$, and the latent factor number is varied in $\{16, 32, 64, 128\}$. The word window size is to 3. For CARL, we select the dimensionality of latent embeddings from $\{15, 50, 100, 200, 300\}$ and the batch size from $\{100, 200\}$. The learning rate is chosen from $\{0.0001, 0.001, 0.01\}$, and the dropout ratio is chosen from $\{0.1, 0.3, 0.5, 0.7, 0.9\}$. The word window size is also set to 3. In NRPA, the dimensionality of embeddings is chosen from $\{16, 32, 64, 128\}$. The number of filters in CNN and the dimension of attention vectors are chosen from $\{20, 40, 80, 100\}$. The word window size is set to 3. For RMG, the number of CNN filters is chosen from $\{50, 100, 150, 200\}$, and the dimensionality of embeddings is chosen from $\{50, 100, 150, 200\}$. The batch size is varied in $\{20, 50, 100, 200\}$, and the dropout ratio is selected from $\{0.1, 0.3, 0.5, 0.7, 0.9\}$. Empirically, the size of the word window is set to 3. For DAML, the dimensionality of latent feature vectors is chosen from $\{8, 16, 32, 64\}$, and the sliding window size is set to 3. The learning rate is selected from $\{0.00001, 0.0001, 0.001, 0.002, 0.01\}$. The dropout ratio is selected from $\{0.1, 0.3, 0.5, 0.7, 0.9\}$.

For all evaluated methods, we employ early stopping strategy for model training, and the optimal hyper-parameters are chosen based on the performances on the validation data.

\subsection{Performance Comparison}

Table~\ref{tab:results} summarizes the results of different recommendation methods. We can make the following observations.

\begin{itemize}
    \item The review-based recommendation methods significantly outperforms PMF, which demonstrates the importance of review data for improving recommendation accuracy.

    \item The neural methods using CNNs and attention mechanism (\eg DeepCoNN and NARRE) achieve better performances than CDL. The potential reason is that CDL employs bag-of-words model to process the review data, thus ignores the word order and local context information.

    \item The review-based methods based on attention mechanism (\eg NARRE, NRPA, RMG) outperform DeepCoNN, by learning the importance of words or reviews for better understanding of users' rating behaviours. DAML usually performs best among baseline methods. It utilizes mutual attention of the CNN to study the correlation between the user reviews and item reviews, thus can capture more information about the user preferences and item properties.

    \item RGNN consistently achieves the best performances on all datasets. The improvements over baselines are statistically significant with $p<0.05$ using student \emph{t-test}. In RGNN, the user/item reviews are described by a directed review graph with preserving the word orders, which can capture the global and non-consecutive topology and dependency information between words in reviews. Moreover, the PGP operator extracts hierarchical and coarser-grained representations of reviews while considering the user/item-specific properties, thus can further help improve the recommendation performance.
\end{itemize}

\begin{figure*}[t]
        \centering
        \begin{subfigure}[b]{0.32\textwidth}
            \centering
            \includegraphics[width=\textwidth]{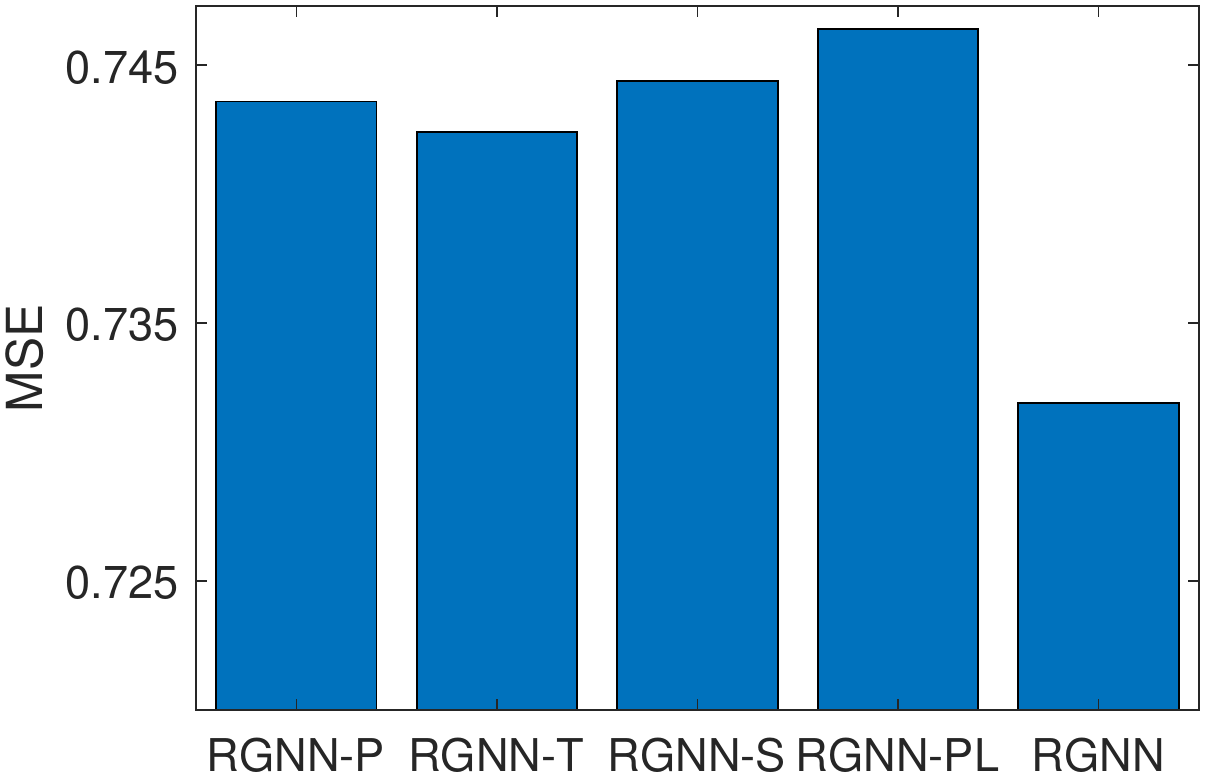}
            \caption[]%
            {Music}
            \label{fig:MusicVar}
        \end{subfigure}
        \hfill
        \begin{subfigure}[b]{0.32\textwidth}
            \centering
            \includegraphics[width=\textwidth]{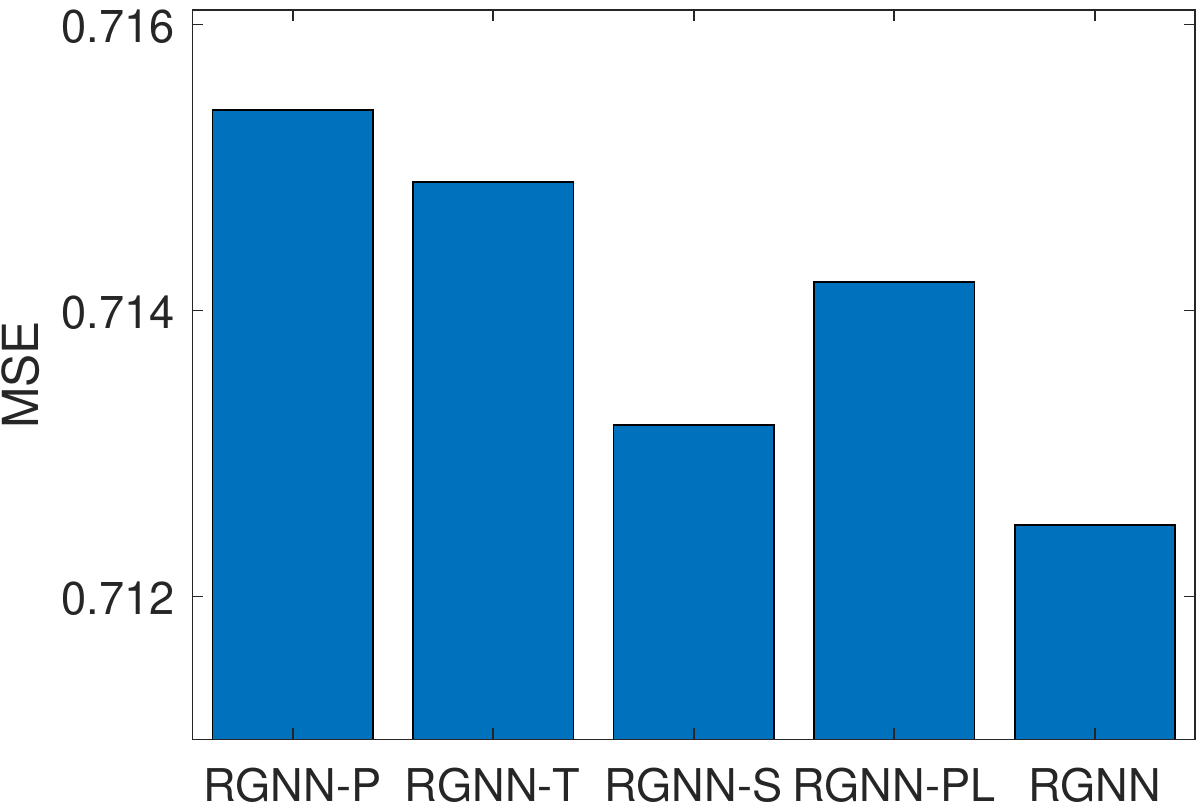}
            \caption[]%
            {Office}
            \label{fig:ToysVar}
        \end{subfigure}
        \hfill
        \begin{subfigure}[b]{0.32\textwidth}
            \centering
            \includegraphics[width=\textwidth]{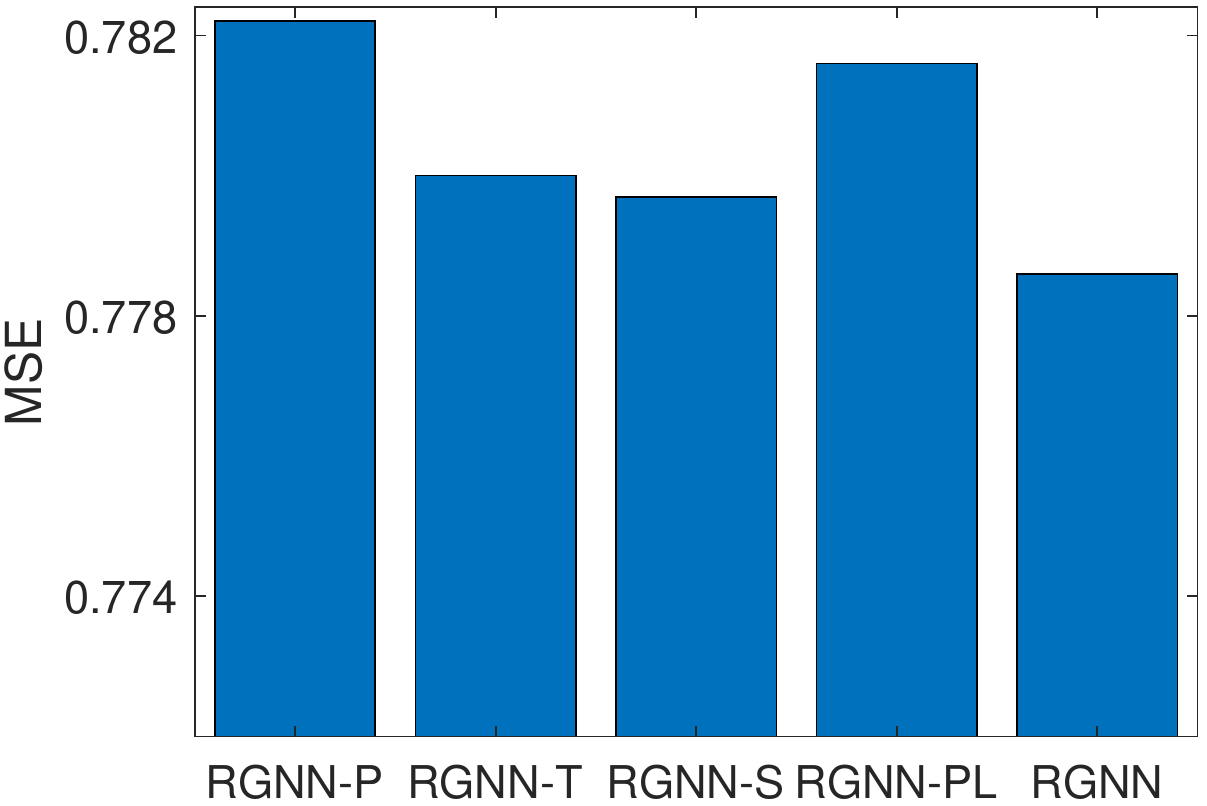}
            \caption[]%
            {Toys}
            \label{fig:ToysVar}
        \end{subfigure}
        \caption{Performances of RGNN-T, RGNN-P, RGNN-S, RGNN-PL, and RGNN on Music, Office, and Toys datasets.}
        \label{fig:variants}
\end{figure*}


Moreover, Table~\ref{tab:LResults} summarizes the performances of RGNN with respect to different settings of $L$. On Music, Office, Toys datasets, the best performances of RGNN are achieved by setting $L$ to 2. On Games, Beauty, and Yelp datasets, RGNN achieves the best performances, when $L$ is set to 3, 4, and 3, respectively. This demonstrates the effectiveness of the hierarchical graph representation learning.

\subsection{Ablation Study}

To investigate the importance of each component of RGNN, we firstly study the contributions of the review graph and the factorization machine layer. We consider the following variants of RGNN for experiments:
\begin{itemize}
    \item \textbf{RGNN-RG}: In this variant, we remove both the user and item review graphs. We use the max pooling of the embeddings of the keywords extracted from $\C{S}_u$ as the review representation of the user $u$, which is denoted by $\B{r}_u$. Then, in Eq.~\eqref{eq:userSem}, we concatenate $\B{m}_u$ with $\B{r}_u$ to form the user representation $\B{p}_u$. Similar operation is performed on the reviews of the item $i$. Note that FM layer is also used for rating prediction in this variant.

    \item \textbf{RGNN-FM}: For this variant, we remove the FM layer and replace the rating prediction function in Eq.~\eqref{eq:fmlayer} by the product of $\B{p}_{u}$ and $\B{q}_i$ as $\hat{y}_{ui}=\B{p}_{u}\B{q}_i^{\top}$.
\end{itemize}
Table~\ref{tab:VariantsGFM} summarizes the performances achieved by DAML, RGNN-RG, RGNN-FM, and RGNN on the Music, Office, and Toys datasets. From Table~\ref{tab:VariantsGFM}, we can note that RGNN outperforms RGNN-RG by 5.01\% and 1.48\% on Music and Toys dataset. Compared with the best baseline method DAML that employs FM layer for prediction, RGNN-FM outperforms DAML on the Office and Toys datasets, achieves comparable results with DAML on the Music dataset. These two observations demonstrate the effectiveness of the proposed review graph based representation learning strategy in capturing the users' preferences on items. Moreover, RGNN outperforms RGNN-FM on Music and Toys datasets. This indicates the FM layer can help improve the rating prediction accuracy. As shown in Table~\ref{tab:VariantsGFM}, RGNN-FM achieves better results than RGNN-RG on three datasets. This demonstrates that the main improvements achieved by RGNN are caused by the proposed hierarchical review graph representation learning strategy.

\begin{figure*}[t]
        \centering
        \begin{subfigure}[b]{0.32\textwidth}
            \centering
            \includegraphics[width=\textwidth]{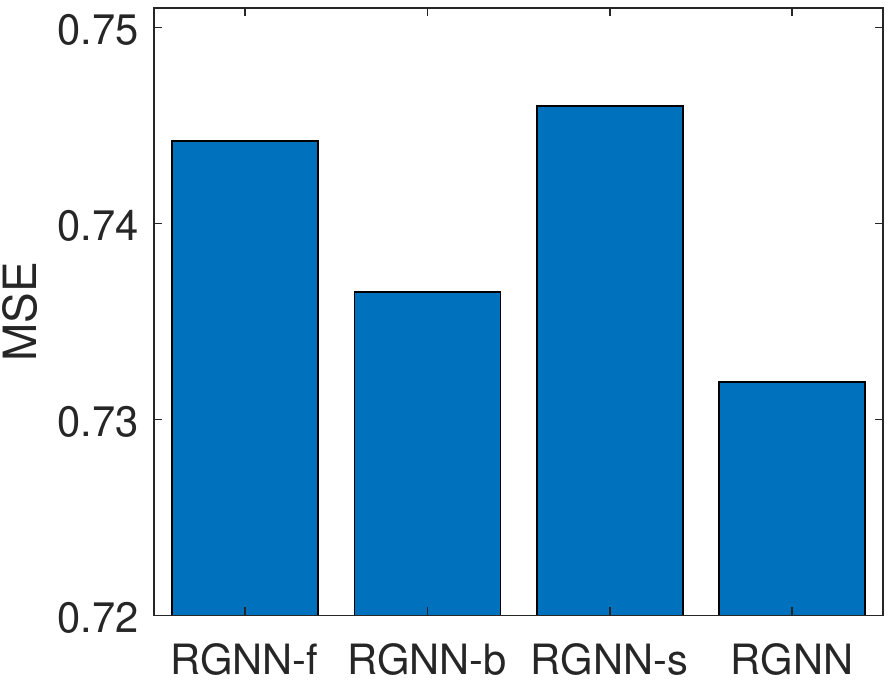}
            \caption[]%
            {Music}
            \label{fig:MusicEdge}
        \end{subfigure}
        \hfill
        \begin{subfigure}[b]{0.32\textwidth}
            \centering
            \includegraphics[width=\textwidth]{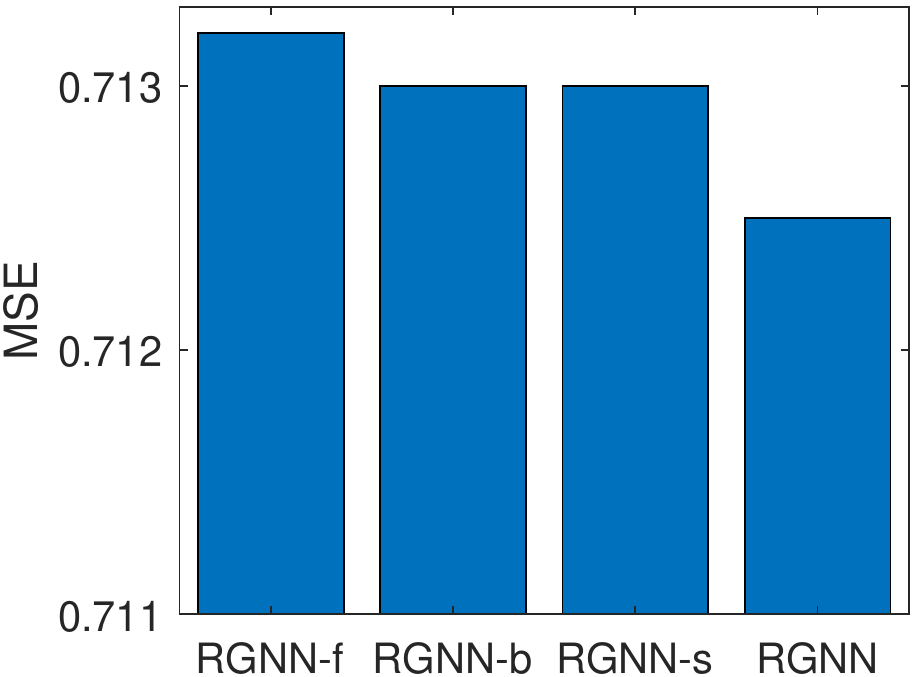}
            \caption[]%
            {Office}
            \label{fig:ToysEdge}
        \end{subfigure}
        \hfill
        \begin{subfigure}[b]{0.32\textwidth}
            \centering
            \includegraphics[width=\textwidth]{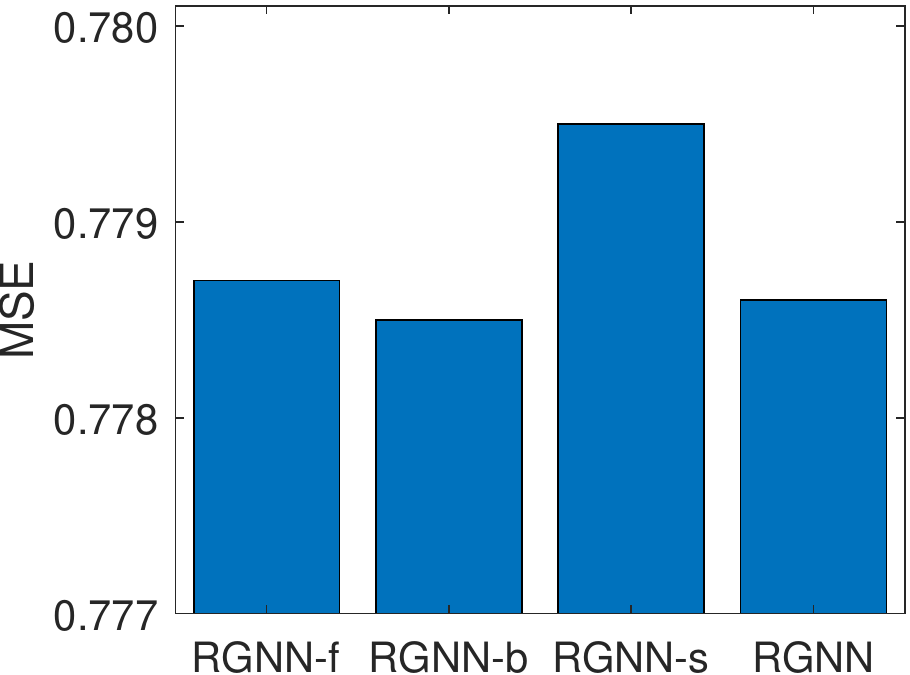}
            \caption[]%
            {Toys}
            \label{fig:ToysEdge}
        \end{subfigure}
        \caption{The performances of RGNN with different settings of edge types on Music, Office, and Toys datasets.}
        \label{fig:EdgeTypes}
\end{figure*}


\begin{figure*}[t]
        \centering
        \begin{subfigure}[b]{0.33\textwidth}
            \centering
            \includegraphics[width=\textwidth]{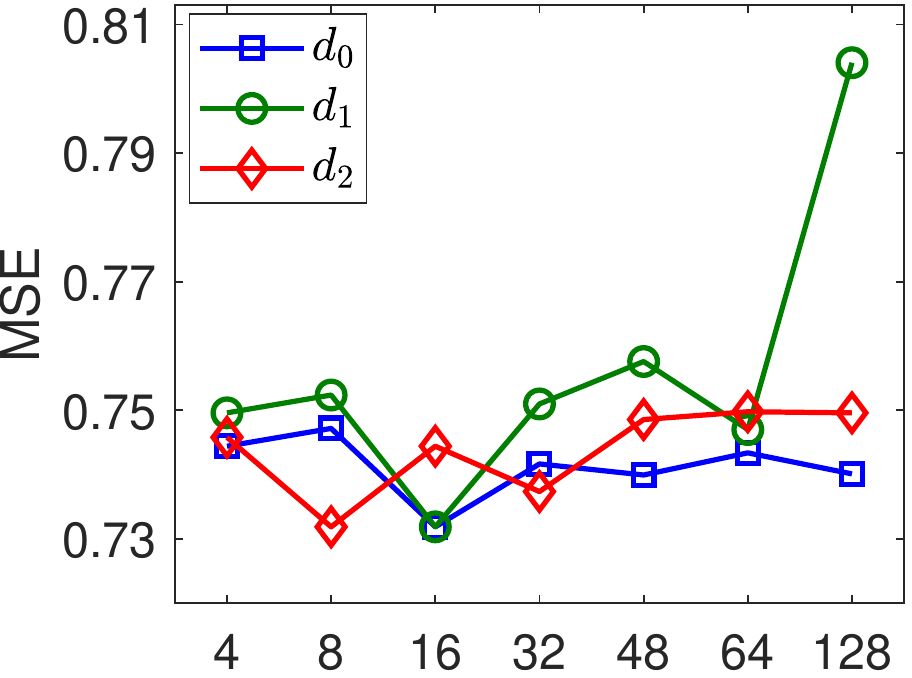}
            \caption[]%
            {$d_0$, $d_1$, and $d_2$}
            \label{fig:d0}
        \end{subfigure}
        \hfill
        \begin{subfigure}[b]{0.33\textwidth}
            \centering
            \includegraphics[width=\textwidth]{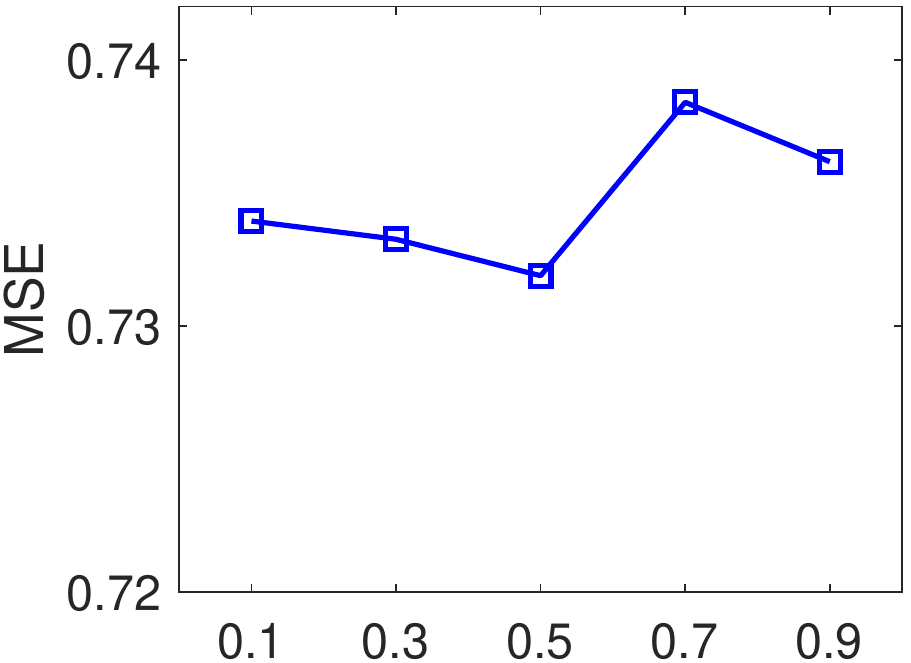}
            \caption[]%
            {$\alpha$}
            \label{fig:alpha}
        \end{subfigure}
        \hfill
        \begin{subfigure}[b]{0.33\textwidth}
            \centering
            \includegraphics[width=\textwidth]{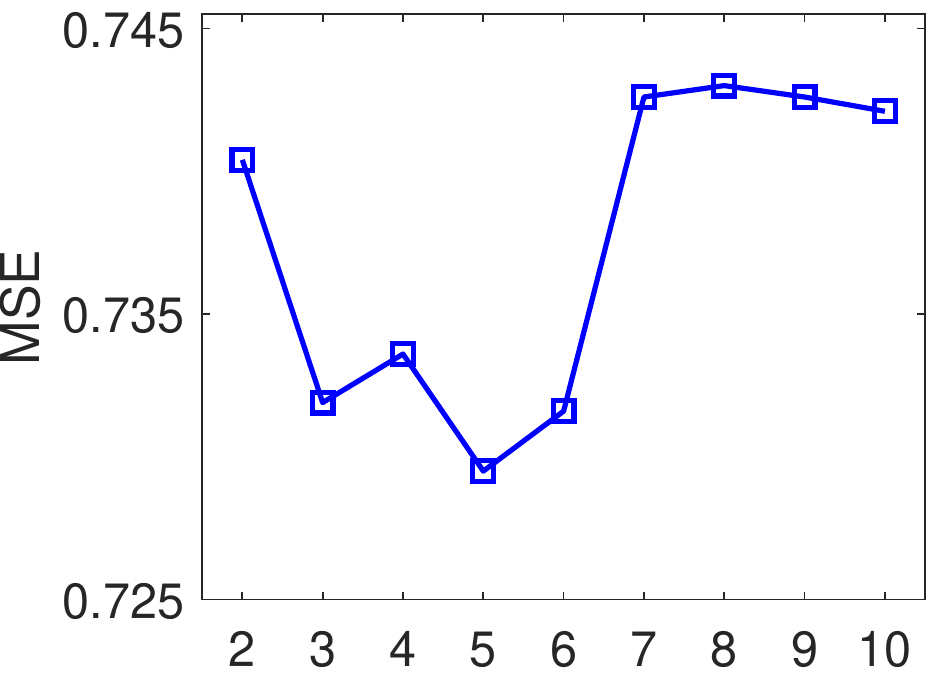}
            \caption[]%
            {$\omega$}
        \end{subfigure}
        \hfill
        \caption{The performances of RGNN with respect to different settings of $d_0$, $d_1$, $d_2$, $\alpha$, and $\omega$ on Music dataset.}
        \label{fig:Hyperparas}
\end{figure*}

\begin{table}[]
\centering
\caption{Performances of DAML, RGNN-RG, RGNN-FM, and RGNN on the Music, Office, and Toys datasets.}
\begin{tabular}{l|cccc}
\hline
Dataset & DAML   & RGNN-RG & RGNN-FM & RGNN   \\ \hline
Music   & 0.7401 & 0.7705  & 0.7411  & \textbf{0.7319} \\ \hline
Office  & 0.7164 & 0.7132  & \textbf{0.7115}  & 0.7125 \\ \hline
Toys    & 0.7909 & 0.7903  & 0.7817  & \textbf{0.7786}\\ \hline
\end{tabular}
\label{tab:VariantsGFM}
\end{table}

Moreover, we perform experiments to analyze the contributions of different components in the hierarchical graph representation learning procedure. Specifically, we evaluate the performances of the following RGNN variants:
\begin{itemize}
  \item \textbf{RGNN-T}: RGNN without considering the word orders (\ie removing the edge type embedding in Eq.~\eqref{eq:attentionscores}).
  \item \textbf{RGNN-P}: RGNN without considering personalized properties of users/items (\ie using the same $\bm{\theta}^\ell$ for all users/items in Eq.~\eqref{eq:infoscore}).
  \item \textbf{RGNN-S}: RGNN without considering the feature diversity of a node's neighborhood in estimating the node importance (\ie removing the second part in Eq.~\eqref{eq:infoscore}).

  \item \textbf{RGNN-PL}: RGNN without considering the graph pooling (\ie removing the PGP operators in Figure~\ref{fig:workflow}).

\end{itemize}
The experimental results on Music, Office, and Toys datasets are summarized in Figure~\ref{fig:variants}. As shown in Figure~\ref{fig:variants}, RGNN outperforms RGNN-T on all three datasets. This indicates that the word order is important in extracting semantic information from reviews for recommendation. Moreover, RGNN achieves better performances than RGNN-P, RGNN-S and RGNN-PL in terms of MSE. This result demonstrates that the proposed PGP operator can help improve the recommendation accuracy via capturing more informative review words for a user/item. When estimating the importance of a node (\ie word), the proposed PGP operator considers the user/item-specific properties, as well as the feature diversity of a node's neighborhood in the review graph. From Figure~\ref{fig:variants}, we can also note that GNN-PL achieves better MSE than RGNN-P on Office and Toys datasets. RGNN-P adopts the same pooling parameter for all users/items. It tends to select a set of common words to form the review representations for all the users/items. The specific properties of a user/item in the review data may not be exploited. Therefore, RGNN-P may perform poorer than RGNN-PL that employs max pooling on the embeddings of all nodes in the review graph of each individual user/item.

\begin{figure*}[t]
\centering
\includegraphics[width=0.95\textwidth]{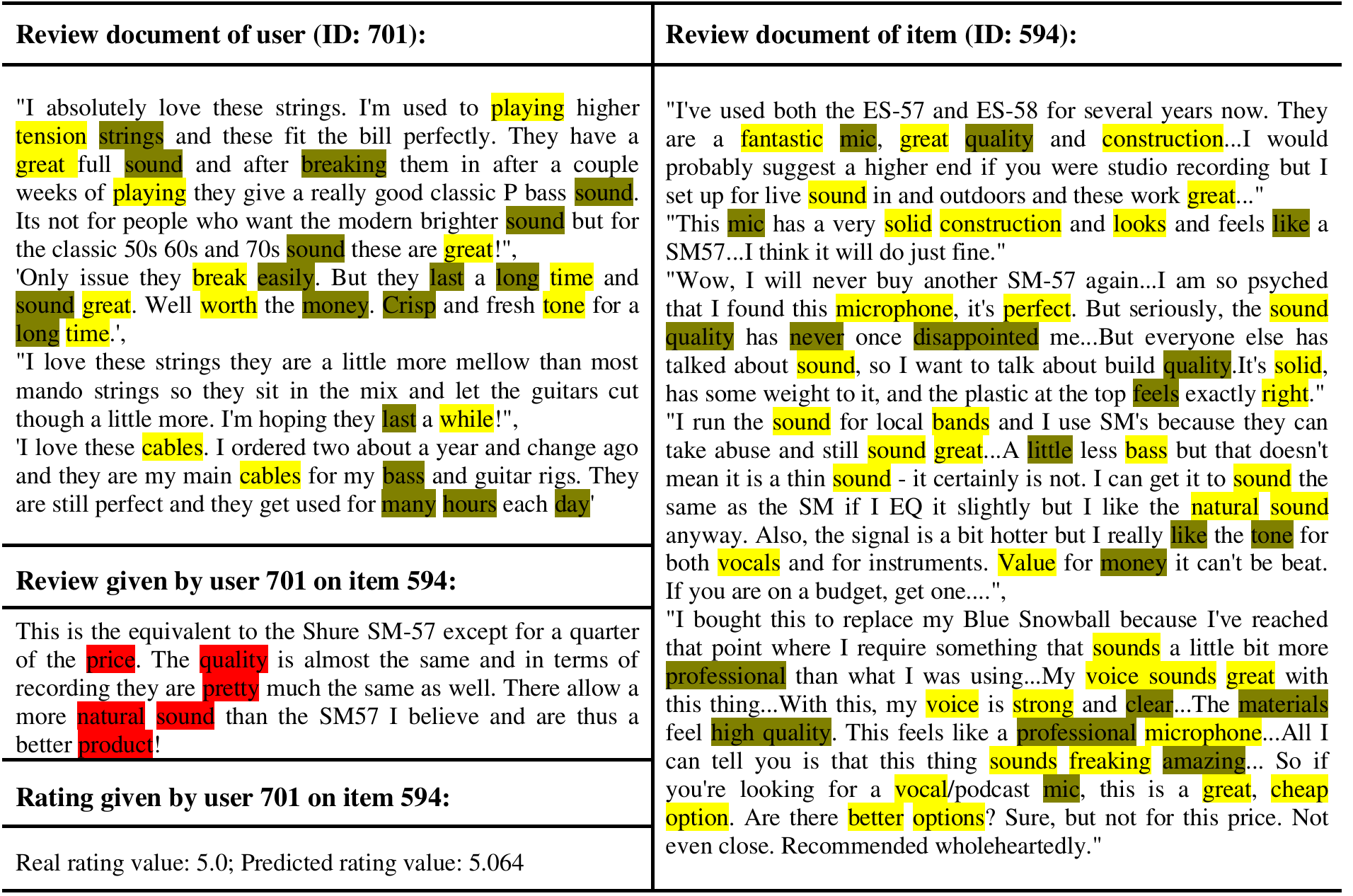}
\caption{Highlighted words according to importance scores in the review documents of a user-item pair. The words with dark yellow background have larger importance scores than the words with yellow background. The words with red background are manually selected to show the user's opinions on the item.}
\label{fig:casestudy}
\end{figure*}

\subsection{Impacts of Edge Types}
In addition, we also conduct experiments to analyze the impacts of different edge types. Specifically, we study the performances of RGNN with the following settings:
\begin{itemize}
  \item \textbf{RGNN-f}: RGNN without considering the forward edge type $e_f$.
  \item \textbf{RGNN-b}: RGNN without considering the backward edge type $e_b$.
  \item \textbf{RGNN-s}: RGNN without considering the self-loop edge type.
\end{itemize}
Figure~\ref{fig:EdgeTypes} summarizes the MSE values achieved by RGNN with different settings of edge types on Music, Office, and Toys datasets. As shown in Figure~\ref{fig:EdgeTypes}, RGNN-b outperforms RGNN-f on all three datasets. This indicates that the forward type of edges are more important than the backward type of edges in capturing users' rating behaviors for recommendation. Moreover, RGNN-s achieves the worst MSE values on Music and Toys datasets. This indicates the self-loop type of edges play an important role in the graph representation learning procedure. This observation is consistent with the definition of the attention mechanism in Eq.~\eqref{eq:attentionvec}. By removing the self-loop type of edges, for each node $x_h$ in the graph at the $\ell$-th layer, it is not included in its first-hop neighbors $\C{N}_h^{\ell}$ in the graph. According to Eq.~\eqref{eq:attentionvec}, the new embedding of $x_h$ (\ie $\bar{\B{x}}_{h}^{\ell}$) is updated by only considering the input embeddings of other neighboring nodes at the $\ell$-th layer, without considering the input embedding of $x_h$ at the $\ell$-th layer. In addition, RGNN achieves the best results on Music and Office datasets. This indicates that the combination of all three types of edges usually provides a better description about the review data, thus achieves more accurate rating prediction results.

\subsection{Parameter Sensitivity Study}

In this section, we perform experiments to evaluate the sensitivity to hyper-parameters for RGNN. The dimensionality of the semantic space $d_0$, the dimensionality of the rating space $d_1$, and the dimensionality of the hidden space of the graph representation learning layers $d_2$, are varied in $\{4, 8, 16, 32, 48, 64, 128\}$. The graph pooling ratio $\alpha$ is adjusted in $\{0.1, 0.3, 0.5, 0.7, 0.9\}$. The size of the word window $\omega$ is chosen from \{2, 3, 4, 5, 6, 7, 8, 9, 10\}. The results on Music dataset are summarized in Figure~\ref{fig:Hyperparas}. We can note that the best MSE can be achieved by setting $d_0$, $d_1$ to 16, and $d_2$ to 8. Larger dimensionality of the latent semantic space and rating space cause more computation cost, however may not help improve the rating prediction accuracy. For the pooling ratio, we can notice that the best setting for $\alpha$ on Music dataset is 0.5. Moreover, RGNN usually achieves better performances by setting $\alpha < 0.5$, compared by setting $\alpha > 0.5$. One potential reason is that when $\alpha > 0.5$ more nodes would be kept by the pooling operation, thus the ``noise nodes'' are more likely to be involved in learning the hierarchical review graph representations. In addition, as shown in Figure~\ref{fig:Hyperparas} (c), RGNN achieves the best MSE value when $\omega$ is set to 5 on Music dataset. Better performances can be usually achieved when $\omega$ is set in the range between 3 and 6.

\subsection{Case Study}
In addition, we also conduct a case study to further explore whether RGNN can select informative words in the graph pooling layer. We sample a user-item pair in test data and extract the review documents of the user and item. Figure~\ref{fig:casestudy} visualizes the heat maps of these review documents. We highlight the words extracted from the first PGP layer of RGNN. 
The words with dark yellow background have larger importance scores than the words with yellow background (refer to Eq.~\eqref{eq:infoscore}).
As shown in Figure~\ref{fig:casestudy}, the words ``playing'', ``sound'', ``money'', and ``long'' are highlighted in user reviews. These words indicate that this user may be more interested in the quality and price of an item. Moreover, in the item reviews, the words ``high'', ``quality'', ``solid'', ``sound'', and ``cheap'' which describe the item properties are selected in the first pooling layer. In the review given by the user on the item, from the words ``price'', ``quality'', ``natural'', and ``sound'', we can argue that the user's high rating on this item is indeed, because this item meets the concerns of the user. Hence, this case again demonstrates that the PGP layer can select informative words from the review graphs, thus can help improve the recommendation accuracy.


\section{Conclusion and Future Work}
\label{sec:conclusion}

In this work, we propose a novel review-based recommendation model, namely Review Graph Neural Network (RGNN), which incorporates the advantages of graph attention networks and hierarchical graph pooling in understanding users' review behaviors for recommendation. In RGNN, the reviews of each user/item are described by a directed graph, which simultaneously considers the co-occurrence of words and the word orders. 
RGNN employs a type-aware graph attention network and a personalized graph pooling operator to learn hierarchical semantic representations for users and items from the review data.
A factorization machine layer is then developed to predict the user's rating on an item, considering the interactions between the user and item semantic features. The experimental results on two real-world datasets show that RGNN consistently outperforms state-of-the-art review-based recommendation methods, in terms of MSE. For future work, we would like to study how to maintain the connectivity of the review graph after the graph pooling operation. Moreover, we are also interested in extending RGNN to solve the top-\emph{N} item recommendation problem.


\bibliographystyle{IEEEtran}

\bibliography{documents}

\begin{thebibliography}{10}
\providecommand{\url}[1]{#1}
\csname url@samestyle\endcsname
\providecommand{\newblock}{\relax}
\providecommand{\bibinfo}[2]{#2}
\providecommand{\BIBentrySTDinterwordspacing}{\spaceskip=0pt\relax}
\providecommand{\BIBentryALTinterwordstretchfactor}{4}
\providecommand{\BIBentryALTinterwordspacing}{\spaceskip=\fontdimen2\font plus
\BIBentryALTinterwordstretchfactor\fontdimen3\font minus
  \fontdimen4\font\relax}
\providecommand{\BIBforeignlanguage}[2]{{%
\expandafter\ifx\csname l@#1\endcsname\relax
\typeout{** WARNING: IEEEtran.bst: No hyphenation pattern has been}%
\typeout{** loaded for the language `#1'. Using the pattern for}%
\typeout{** the default language instead.}%
\else
\language=\csname l@#1\endcsname
\fi
#2}}
\providecommand{\BIBdecl}{\relax}
\BIBdecl

\bibitem{shi2014collaborative}
Y.~Shi, M.~Larson, and A.~Hanjalic, ``Collaborative filtering beyond the
  user-item matrix: A survey of the state of the art and future challenges,''
  \emph{CSUR}, vol.~47, no.~1, p.~3, 2014.

\bibitem{zhang2019deep}
S.~Zhang, L.~Yao, A.~Sun, and Y.~Tay, ``Deep learning based recommender system:
  A survey and new perspectives,'' \emph{CSUR}, vol.~52, no.~1, p.~5, 2019.

\bibitem{hu2014your}
L.~Hu, A.~Sun, and Y.~Liu, ``Your neighbors affect your ratings: on
  geographical neighborhood influence to rating prediction,'' in
  \emph{Proceedings of the 37th international ACM SIGIR conference on Research
  \& development in information retrieval}, 2014, pp. 345--354.

\bibitem{liu2013personalized}
X.~Liu, Y.~Liu, K.~Aberer, and C.~Miao, ``Personalized point-of-interest
  recommendation by mining users' preference transition,'' in \emph{Proceedings
  of the 22nd ACM international conference on Information \& Knowledge
  Management}, 2013, pp. 733--738.

\bibitem{wu2019pd}
Q.~Wu, Y.~Liu, C.~Miao, B.~Zhao, Y.~Zhao, and L.~Guan, ``Pd-gan: adversarial
  learning for personalized diversity-promoting recommendation,'' in
  \emph{IJCAI'19}.\hskip 1em plus 0.5em minus 0.4em\relax AAAI Press, 2019, pp.
  3870--3876.

\bibitem{sun2018recurrent}
Z.~Sun, J.~Yang, J.~Zhang, A.~Bozzon, L.-K. Huang, and C.~Xu, ``Recurrent
  knowledge graph embedding for effective recommendation,'' in
  \emph{RecSys'18}.\hskip 1em plus 0.5em minus 0.4em\relax ACM, 2018, pp.
  297--305.

\bibitem{chen2015recommender}
L.~Chen, G.~Chen, and F.~Wang, ``Recommender systems based on user reviews: the
  state of the art,'' \emph{User Modeling and User-Adapted Interaction},
  vol.~25, no.~2, pp. 99--154, 2015.

\bibitem{blei2003latent}
D.~M. Blei, A.~Y. Ng, and M.~I. Jordan, ``Latent dirichlet allocation,''
  \emph{JMLR}, vol.~3, no. Jan, pp. 993--1022, 2003.

\bibitem{mikolov2013efficient}
T.~Mikolov, K.~Chen, G.~Corrado, and J.~Dean, ``Efficient estimation of word
  representations in vector space,'' \emph{arXiv preprint arXiv:1301.3781},
  2013.

\bibitem{mcauley2013hidden}
J.~McAuley and J.~Leskovec, ``Hidden factors and hidden topics: understanding
  rating dimensions with review text,'' in \emph{RecSys'13}, 2013, pp.
  165--172.

\bibitem{bao2014topicmf}
Y.~Bao, H.~Fang, and J.~Zhang, ``Topicmf: Simultaneously exploiting ratings and
  reviews for recommendation,'' in \emph{AAAI'14}, 2014.

\bibitem{ling2014ratings}
G.~Ling, M.~R. Lyu, and I.~King, ``Ratings meet reviews, a combined approach to
  recommend,'' in \emph{RecSys'14}, 2014, pp. 105--112.

\bibitem{tan2016rating}
Y.~Tan, M.~Zhang, Y.~Liu, and S.~Ma, ``Rating-boosted latent topics:
  Understanding users and items with ratings and reviews.'' in \emph{IJCAI'16},
  vol.~16, 2016, pp. 2640--2646.

\bibitem{zhang2016collaborative}
W.~Zhang, Q.~Yuan, J.~Han, and J.~Wang, ``Collaborative multi-level embedding
  learning from reviews for rating prediction.'' in \emph{IJCAI'16}, 2016, pp.
  2986--2992.

\bibitem{wang2015collaborative}
H.~Wang, N.~Wang, and D.-Y. Yeung, ``Collaborative deep learning for
  recommender systems,'' in \emph{KDD'15}.\hskip 1em plus 0.5em minus
  0.4em\relax ACM, 2015, pp. 1235--1244.

\bibitem{zheng2017joint}
L.~Zheng, V.~Noroozi, and P.~S. Yu, ``Joint deep modeling of users and items
  using reviews for recommendation,'' in \emph{WSDM'17}.\hskip 1em plus 0.5em
  minus 0.4em\relax ACM, 2017, pp. 425--434.

\bibitem{chen2018neural}
C.~Chen, M.~Zhang, Y.~Liu, and S.~Ma, ``Neural attentional rating regression
  with review-level explanations,'' in \emph{WWW'18}, 2018, pp. 1583--1592.

\bibitem{liu2019daml}
D.~Liu, J.~Li, B.~Du, J.~Chang, and R.~Gao, ``Daml: Dual attention mutual
  learning between ratings and reviews for item recommendation,'' in
  \emph{KDD'19}.\hskip 1em plus 0.5em minus 0.4em\relax ACM, 2019, pp.
  344--352.

\bibitem{liu2019nrpa}
H.~Liu, F.~Wu, W.~Wang, X.~Wang, P.~Jiao, C.~Wu, and X.~Xie, ``Nrpa: Neural
  recommendation with personalized attention,'' \emph{SIGIR'19}, 2019.

\bibitem{bauman2017aspect}
K.~Bauman, B.~Liu, and A.~Tuzhilin, ``Aspect based recommendations:
  Recommending items with the most valuable aspects based on user reviews,'' in
  \emph{Proceedings of the 23rd ACM SIGKDD International Conference on
  Knowledge Discovery and Data Mining}, 2017, pp. 717--725.

\bibitem{chin2018anr}
J.~Y. Chin, K.~Zhao, S.~Joty, and G.~Cong, ``Anr: Aspect-based neural
  recommender,'' in \emph{Proceedings of the 27th ACM International Conference
  on Information and Knowledge Management}, 2018, pp. 147--156.

\bibitem{wu2019context}
L.~Wu, C.~Quan, C.~Li, Q.~Wang, B.~Zheng, and X.~Luo, ``A context-aware
  user-item representation learning for item recommendation,'' \emph{TOIS},
  vol.~37, no.~2, p.~22, 2019.

\bibitem{lu2018coevolutionary}
Y.~Lu, R.~Dong, and B.~Smyth, ``Coevolutionary recommendation model: Mutual
  learning between ratings and reviews,'' in \emph{WWW'18}, 2018, pp. 773--782.

\bibitem{wu2019reviews}
C.~Wu, F.~Wu, T.~Qi, S.~Ge, Y.~Huang, and X.~Xie, ``Reviews meet graphs:
  Enhancing user and item representations for recommendation with hierarchical
  attentive graph neural network,'' in \emph{EMNLP-IJCNLP 2019}, 2019, pp.
  4886--4895.

\bibitem{wu2020comprehensive}
Z.~Wu, S.~Pan, F.~Chen, G.~Long, C.~Zhang, and S.~Y. Philip, ``A comprehensive
  survey on graph neural networks,'' \emph{IEEE Transactions on Neural Networks
  and Learning Systems}, 2020.

\bibitem{scarselli2008graph}
F.~Scarselli, M.~Gori, A.~C. Tsoi, M.~Hagenbuchner, and G.~Monfardini, ``The
  graph neural network model,'' \emph{IEEE Transactions on Neural Networks},
  vol.~20, no.~1, pp. 61--80, 2008.

\bibitem{gallicchio2010graph}
C.~Gallicchio and A.~Micheli, ``Graph echo state networks,'' in \emph{The 2010
  International Joint Conference on Neural Networks (IJCNN)}.\hskip 1em plus
  0.5em minus 0.4em\relax IEEE, 2010, pp. 1--8.

\bibitem{Li2015Gated}
Y.~Li, D.~Tarlow, M.~Brockschmidt, and R.~Zemel, ``Gated graph sequence neural
  networks,'' in \emph{ICLR}, 2015.

\bibitem{Bruna2013Spectral}
J.~Bruna, W.~Zaremba, A.~Szlam, and Y.~Lecun, ``Spectral networks and locally
  connected networks on graphs,'' in \emph{ICLR}, 2014.

\bibitem{deffer2016gcnn}
M.~Defferrard, X.~Bresson, and P.~Vandergheynst, ``Convolutional neural
  networks on graphs with fast localized spectral filtering,'' in \emph{NIPS},
  2016, pp. 3844--3852.

\bibitem{kipf2016semi}
T.~N. Kipf and M.~Welling, ``Semi-supervised classification with graph
  convolutional networks,'' \emph{arXiv preprint arXiv:1609.02907}, 2016.

\bibitem{micheli2009neural}
A.~Micheli, ``Neural network for graphs: a contextual constructive approach,''
  \emph{IEEE Transactions on Neural Networks}, vol.~20, no.~3, pp. 498--511,
  2009.

\bibitem{atwood2016dcnn}
J.~Atwood and D.~Towsley, ``Diffusion-convolutional neural networks,'' in
  \emph{Advances in Neural Information Processing Systems 29}, 2016, pp.
  1993--2001.

\bibitem{hamilton2017gsage}
W.~Hamilton, Z.~Ying, and J.~Leskovec, ``Inductive representation learning on
  large graphs,'' in \emph{Advances in Neural Information Processing Systems
  30}, 2017, pp. 1024--1034.

\bibitem{velivckovic2017graph}
P.~Veli{\v{c}}kovi{\'c}, G.~Cucurull, A.~Casanova, A.~Romero, P.~Lio, and
  Y.~Bengio, ``Graph attention networks,'' \emph{arXiv preprint
  arXiv:1710.10903}, 2017.

\bibitem{Zhang2018GaAN}
J.~Zhang, X.~Shi, J.~Xie, H.~Ma, I.~King, and D.~Y. Yeung, ``Gaan: Gated
  attention networks for learning on large and spatiotemporal graphs,'' in
  \emph{UAI}, 2018.

\bibitem{keyu2019GIN}
J.~L. Keyulu~Xu, Weihua~Hu and S.~Jegelka, ``How powerful are graph neural
  networks?'' in \emph{ICLR}, 2019.

\bibitem{defferrard2016convolutional}
M.~Defferrard, X.~Bresson, and P.~Vandergheynst, ``Convolutional neural
  networks on graphs with fast localized spectral filtering,'' in
  \emph{Advances in neural information processing systems}, 2016, pp.
  3844--3852.

\bibitem{fey2018splinecnn}
M.~Fey, J.~Eric~Lenssen, F.~Weichert, and H.~M{\"u}ller, ``Splinecnn: Fast
  geometric deep learning with continuous b-spline kernels,'' in
  \emph{Proceedings of the IEEE Conference on Computer Vision and Pattern
  Recognition}, 2018, pp. 869--877.

\bibitem{bianchi2019mincut}
F.~M. Bianchi, D.~Grattarola, and C.~Alippi, ``Mincut pooling in graph neural
  networks,'' \emph{arXiv preprint arXiv:1907.00481}, 2019.

\bibitem{ma2019graph}
Y.~Ma, S.~Wang, C.~C. Aggarwal, and J.~Tang, ``Graph convolutional networks
  with eigenpooling,'' \emph{arXiv preprint arXiv:1904.13107}, 2019.

\bibitem{vinyals2015order}
O.~Vinyals, S.~Bengio, and M.~Kudlur, ``Order matters: Sequence to sequence for
  sets,'' \emph{arXiv preprint arXiv:1511.06391}, 2015.

\bibitem{zhang2018end}
M.~Zhang, Z.~Cui, M.~Neumann, and Y.~Chen, ``An end-to-end deep learning
  architecture for graph classification,'' in \emph{Thirty-Second AAAI
  Conference on Artificial Intelligence}, 2018.

\bibitem{ying2018hierarchical}
Z.~Ying, J.~You, C.~Morris, X.~Ren, W.~Hamilton, and J.~Leskovec,
  ``Hierarchical graph representation learning with differentiable pooling,''
  in \emph{Advances in Neural Information Processing Systems}, 2018, pp.
  4800--4810.

\bibitem{gao2019graph}
H.~Gao and S.~Ji, ``Graph u-nets,'' \emph{arXiv preprint arXiv:1905.05178},
  2019.

\bibitem{gao2019learning}
H.~Gao, Y.~Chen, and S.~Ji, ``Learning graph pooling and hybrid convolutional
  operations for text representations,'' in \emph{WWW'19}.\hskip 1em plus 0.5em
  minus 0.4em\relax ACM, 2019, pp. 2743--2749.

\bibitem{lee2019self}
J.~Lee, I.~Lee, and J.~Kang, ``Self-attention graph pooling,'' \emph{arXiv
  preprint arXiv:1904.08082}, 2019.

\bibitem{diehl2019edge}
F.~Diehl, ``Edge contraction pooling for graph neural networks,'' \emph{arXiv
  preprint arXiv:1905.10990}, 2019.

\bibitem{ranjan2019asap}
E.~Ranjan, S.~Sanyal, and P.~P. Talukdar, ``Asap: Adaptive structure aware
  pooling for learning hierarchical graph representations,'' \emph{arXiv
  preprint arXiv:1911.07979}, 2019.

\bibitem{zhang2019hierarchical}
Z.~Zhang, J.~Bu, M.~Ester, J.~Zhang, C.~Yao, Z.~Yu, and C.~Wang, ``Hierarchical
  graph pooling with structure learning,'' \emph{arXiv preprint
  arXiv:1911.05954}, 2019.

\bibitem{rendle2010factorization}
S.~Rendle, ``Factorization machines,'' in \emph{ICDM'10}, 2010, pp. 995--1000.

\bibitem{malliaros2017graph}
F.~D. Malliaros and M.~Vazirgiannis, ``Graph-based text representations:
  Boosting text mining, nlp and information retrieval with graphs,'' in
  \emph{EMNLP'17: Tutorial Abstracts}, 2017.

\bibitem{mihalcea-tarau-2004-textrank}
R.~Mihalcea and P.~Tarau, ``{T}ext{R}ank: Bringing order into text,'' in
  \emph{EMNLP'04}, Jul. 2004, pp. 404--411.

\bibitem{shen2018baseline}
D.~Shen, G.~Wang, W.~Wang, M.~R. Min, Q.~Su, Y.~Zhang, C.~Li, R.~Henao, and
  L.~Carin, ``Baseline needs more love: On simple word-embedding-based models
  and associated pooling mechanisms,'' in \emph{ACL'18}, 2018, pp. 440--450.

\bibitem{he2016ups}
R.~He and J.~McAuley, ``Ups and downs: Modeling the visual evolution of fashion
  trends with one-class collaborative filtering,'' in \emph{WWW'16}, 2016, pp.
  507--517.

\bibitem{mnih2008probabilistic}
A.~Mnih and R.~R. Salakhutdinov, ``Probabilistic matrix factorization,'' in
  \emph{NIPS'08}, 2008, pp. 1257--1264.

\bibitem{koren2008factorization}
Y.~Koren, ``Factorization meets the neighborhood: a multifaceted collaborative
  filtering model,'' in \emph{Proceedings of the 14th ACM SIGKDD international
  conference on Knowledge discovery and data mining}, 2008, pp. 426--434.

\bibitem{kingma2014adam}
D.~P. Kingma and J.~Ba, ``Adam: A method for stochastic optimization,''
  \emph{arXiv preprint arXiv:1412.6980}, 2014.

\end{thebibliography}

\end{document}